\documentclass[sigconf]{acmart}
 
\usepackage{xcolor}
\usepackage{fontawesome}
\usepackage{soul}
\usepackage{textcomp}
\usepackage{mathtools}

\newcommand{\edit}[1]{{\textcolor{black}{#1}}}
\usepackage{xspace}
\newcommand{\system}{Spellburst\xspace} 

\usepackage{balance}
\usepackage{listings}
\usepackage{float}

\definecolor{antiquewhite}{rgb}{0.98, 0.92, 0.84}
\definecolor{beige}{rgb}{0.96, 0.96, 0.86}
\definecolor{bubbles}{rgb}{0.91, 1.0, 1.0}
\definecolor{ivory}{rgb}{1.0, 1.0, 0.94}
\definecolor{modify}{HTML}{003F5C}
\definecolor{sketch}{HTML}{7A5195}
\definecolor{merge}{HTML}{EF5675} 

\definecolor{lightgray}{rgb}{.9,.9,.9}
\definecolor{darkgray}{rgb}{.4,.4,.4}
\definecolor{purple}{rgb}{0.65, 0.12, 0.82}

\lstdefinelanguage{JavaScript}{
  keywords={typeof, new, true, false, catch, function, return, null, catch, switch, var, if, in, while, do, else, case, break},
  keywordstyle=\color{blue}\bfseries,
  ndkeywords={class, export, boolean, throw, implements, import, this},
  ndkeywordstyle=\color{darkgray}\bfseries,
  identifierstyle=\color{black},
  sensitive=false,
  comment=[l]{//},
  morecomment=[s]{/*}{*/},
  commentstyle=\color{purple}\ttfamily,
  stringstyle=\color{red}\ttfamily,
  morestring=[b]',
  morestring=[b]"
}

\usepackage{lipsum}

\newcommand\blfootnote[1]{%
  \begingroup
  \renewcommand\thefootnote{}\footnote{#1}%
  \addtocounter{footnote}{-1}%
  \endgroup
}

\AtBeginDocument{%
  \providecommand\BibTeX{{%
    \normalfont B\kern-0.5em{\scshape i\kern-0.25em b}\kern-0.8em\TeX}}}

\copyrightyear{2023}
\acmYear{2023}
\setcopyright{acmlicensed}\acmConference[UIST '23]{The 36th Annual ACM Symposium on User Interface Software and Technology}{October 29-November 1, 2023}{San Francisco, CA, USA}
\acmBooktitle{The 36th Annual ACM Symposium on User Interface Software and Technology (UIST '23), October 29-November 1, 2023, San Francisco, CA, USA}
\acmPrice{15.00}
\acmDOI{10.1145/3586183.3606719}
\acmISBN{979-8-4007-0132-0/23/10}
\begin{document}


\title[\system: Creative Coding with LLMs]{\system: A Node-based Interface for Exploratory Creative Coding with Natural Language Prompts}

\author{Tyler Angert\textsuperscript{*}}
 \email{tyler@replit.com}
\affiliation{%
  \institution{Replit}
  \city{San Francisco}
  \country{USA}
}

\author{Miroslav Ivan Suzara\textsuperscript{*}}
 \email{msuzara@stanford.edu}
\affiliation{%
  \institution{Stanford University}
  \city{Stanford}
  \country{USA}
}

\author{Jenny Han\textsuperscript{*}}
 \email{jennyhan@cs.stanford.edu}
\affiliation{%
  \institution{Stanford University}
  \city{Stanford}
  \country{USA}
}

\author{Christopher Lawrence Pondoc}
 \email{clpondoc@stanford.edu}
\affiliation{%
  \institution{Stanford University}
  \city{Stanford}
  \country{USA}
}

\author{Hariharan Subramonyam}
 \email{harihars@stanford.edu}
\affiliation{%
  \institution{Stanford University}
  \city{Stanford}
  \country{USA}
}

\renewcommand{\shortauthors}{Angert et al.}
\begin{abstract} 
Creative coding tasks are often exploratory in nature. When producing digital artwork, artists usually begin with a high-level semantic construct such as a ``stained glass filter'' and programmatically implement it by varying code parameters such as shape, color, lines, and opacity to produce visually appealing results. Based on interviews with artists, it can be effortful to translate semantic constructs to program syntax, and current programming tools don't lend well to rapid creative exploration. To address these challenges, we introduce \system, a large language model (LLM) powered creative-coding environment. \system provides (1) a node-based interface that allows artists to create generative art and explore variations through branching and merging operations, (2) expressive prompt-based interactions to engage in semantic programming, and (3) dynamic prompt-driven interfaces and direct code editing to seamlessly switch between semantic and syntactic exploration. Our evaluation with artists demonstrates \system's potential to enhance creative coding practices and inform the design of computational creativity tools that bridge semantic and syntactic spaces.
\blfootnote{\textsuperscript{*} indicates equal contribution by authors.}
\end{abstract}


\begin{teaserfigure}
  \centering
   \includegraphics[width=0.9\textwidth]{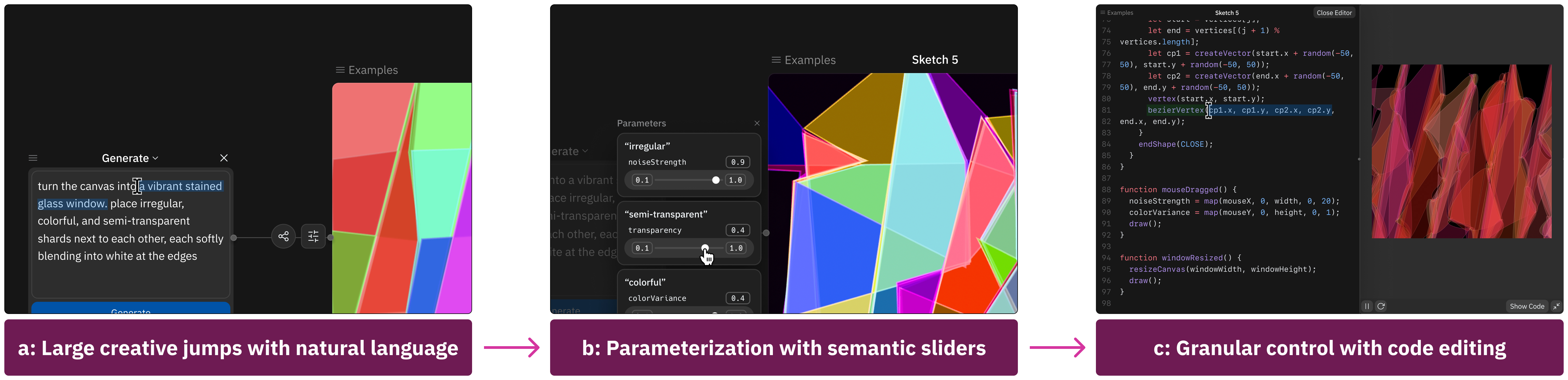}
 \caption{Exploratory Creative Coding using \system: (a) the artist uses a natural language prompt to generate an initial sketch based on semantic intent, (b) uses the dynamic prompt-driven slides to rapidly refine the output, and (c) performs fine-grained adjustments through direct code-editing.}
  \label{fig:teaser}
  \Description[]{}
\end{teaserfigure}



\maketitle
\section{Introduction}

Creative coding is an \textit{expressive} and \textit{exploratory} activity. When producing artwork, artists often begin with high-level constructs grounded in mimesis or metaphors and programmatically explore digital renditions of their ideas~\cite{gaut2003creativity, halliwell2009aesthetics, knochel2015if}. For instance, imagine creating art inspired by the reflective behavior of light. Starting with the idea of stained glass windows, the artist may write code to generate a glass effect on the digital canvas. They may then rapidly explore variations by adjusting the color and translucency of the glass, angle of light reflection, amount of refraction, etc. Finally, based on outcomes, they may take a larger creative jump and explore a different construct, such as the reflection of moonlight on water. Across these exploratory tasks, artists (1) repeatedly translate expressive semantic intents into lines of code, (2) keep track of different creative variations, and (3) compare, combine, and extend emergent outputs to further exploration. Unfortunately, existing programming environments don't lend themselves well to this form of creative exploratory process~\cite{exploringexploratory}. While the needs of artists engaged in creative coding tasks overlap with traditional programming, they require specific affordances for rapid iteration and semantic exploration. 

First, creative coding is experimentation-heavy. During exploration, artists would benefit from systems that help them quickly and intentionally keep track of intermediate visual outputs and corresponding specifications. Current versioning strategies are largely disconnected from programming environments and demand a separate workflow to commit and retrieve program states. This introduces friction in artists' creative processes to document outputs for later consideration. Artists may have to repeatedly capture screenshots of their evolving art and manually associate them with corresponding code either through versioned code files or commented code snippets. Second, given the emphasis on \textit{visual} outputs as feedback and inspiration for subsequent exploration, artists would benefit from ways to visualize and compare generated outputs. Current programming tool layouts primarily consist of a side-by-side text editor and output view, making it challenging to visually compare results across multiple ``runs'' of the program. While recent tools address issues of comparing different outputs~\cite{zaman2015gem, 10.1145/3290605.3300758, kim2022mixplorer}, they are still primarily disconnected from generative coding tasks (e.g., combining line attributes from one output with color from a different exploratory path)~\cite{boden2010creativity}. 

Third, as mentioned in the light reflection example, larger creative jumps are often grounded in semantic constructs. When such jumps occur during creative thinking, artists spend a significant amount of time programmatically authoring initial representations of the new idea. Consequently, they may take fewer creative shifts and fixate on incremental variations of their initial ideas~\cite{nijstad2002cognitive,kleinmintz2019two}. An advantage of the recent generative AI tools based on large language models (LLMs) is that they allow artists to quickly explore divergent ideas through expressive natural language prompts (e.g., Adobe Firefly~\cite{adobefirefly}). However, finer editing using natural language prompts can be tedious (e.g., ``increase the line thickness by 5 pixels''). Artists benefit from executing larger creative shifts using semantic prompts, while program syntax (or dedicated interface controls) may be better suited for fine-grained exploration and control. Thus, the motivating question for our work is \textit{how can we incorporate generative AI capabilities within current creative programming workflows to support exploratory generative art-making?}

By conducting formative interviews with ten expert creative coders, we identified ways in which artists navigated semantic and syntactic spaces, distinct tasks in their creative exploratory workflows, and pain points due to limitations of existing tools. Based on insights from the study, we developed \textit{\system}, a visual interface for exploratory programming. The primary view in \system is a \textit{node-based} layout in which each node represents a creative output along with backing code. Artists can use branching and merging interactions to clone, extend, and combine different variations. As shown in Figure~\ref{fig:teaser}, to execute larger semantic jumps, \system supports a natural language \textit{prompting interface} to generate code, and artists can further refine the generated code either through dynamic prompt-driven interface controls or directly edit the code in the code editor. We conducted an online survey with 52 crowd-sourced participants to understand the range of natural language prompts for specific creative tasks. Based on insights from the survey, we implemented \system to support seamless switching between prompt-based exploration and program editing, and the visual layout helps them keep track of the creative exploration process. We evaluated \system with 10 expert artists across controlled and open-ended creative tasks.

We contribute (1) \system, a node-based visual programming interface for exploratory programming tasks, (2) expressive prompting support for creative experimentation, and (3) interface affordances for semantic-syntactic integration to flexibly execute larger creative shifts and fine-grained code-editing. 
\section{Related Work}

\textit{Creative coding} is a branch of computer programming where the purpose is to create \textit{artistic} and \textit{expressive} output rather than functional output~\cite{shiffman2012nature, peppler2005creative}. \textit{Generative} art is a specific approach to creative coding in which artists \textit{``program computers to undertake creative instructions''} ~\cite{ward1999drew, drawingwithcode, tempel2017generative}. Mathematical and computational algorithms are central to this approach. In producing art, artists engage in the exploration and experimentation of code to produce variations of algorithms and outputs~\cite{exploringexploratory, kery2017variolite, kerydesigning}. During this process, they encounter phenomena such as emergence, randomness, and interaction to produce creative results~\cite{thecreatorsproject_2012, ward1999drew, drawingwithcode, thealgorists}.

Today, artists may use programming tools such as Processing \cite{processing}, p5.js \cite{mccarthy2015getting}, OpenFrameworks \cite{lieberman2009openframeworks}, and TouchDesigner~\cite{touchdesigner} to create their artwork. Within these tools, exploration by programming takes place through \textit{edit-run cycles} in close proximity \cite{kery2017variolite, microversioning, kerydesigning}. However, this can also vary by scale (i.e., changing a specific variable versus file-level changes) and duration (i.e., pertaining to one particular element versus a broader computational model that is being explored) \cite{exploringexploratory}. This contrasts with traditional conceptualizations of programming, where specifications are typically mapped out in advance and for functional purposes. In \system, our goal is to support open-ended creative experimentation. 

We narrow our focus to three specific needs for creativity support tools (CST) within the realm of exploratory creative coding: (1) version control systems for \textit{history-keeping}~\cite{shneiderman2007creativity}, (2) higher \textit{closeness of mapping} between semantic concepts and syntactic representations in creativity support tools~\cite{exploringexploratory, green1996cognitivedimensions}, and (3) AI augmentation for creative exploration~\cite{shneiderman2007creativity, frich2019mapping}.

\subsection{Tracking Exploration History}
A user's personal code history is an important site for exploration and for management of alternatives, especially in creative settings~\cite{sterman2022towards, shneiderman2007creativity}. Existing history-keeping or version control systems such as Git offer features for documenting previous alternatives but are costly to use in terms of time and effort~\cite{kery2017variolite, exploringexploratory, stamper}. Creative coders might prefer informal or manual versioning methods such as taking screenshots or screen captures of each run and copy-pasting code~\cite{kery2017variolite,li2021we, sterman2022towards}. Second, current version control systems are not explicitly designed for rapid iteration and remixing on prior states. Creative work is often non-linear; artists jump back to previous work or explore multiple ideas in parallel~\cite{li2021we}. Using tools such as Git to create new branches, revert to previous states, or merge branches creates friction that disrupts an artist's flow or process. 

Previous qualitative studies with artists emphasize the importance of designing and co-designing with artists' mental models for code and code versions in mind~\cite{li2021we, sterman2022towards, rawn}. Like artists who work with physical media, generative artists also engage deeply with the materiality of their work~\cite{rawn}; they pay special attention to the properties, expressiveness, and craftsmanship of their software~\cite{rawn, immaterialmaterial_2010}. They may view their versions as panels to ``pin up'' for review \cite{stamper}, or they may think of their versions as a ``palette of materials'' with which to mix, experiment, and use \cite{sterman2022towards}. 

Existing work on novel version control systems for coding has aimed to support rapid iteration and exploration by allowing parallel source editing and execution \cite{2008design}, providing a sandbox environment to quickly test, save or discard ``microversions'' of code \cite{kery2017variolite, microversioning}, and storing each commit as a node in a graph alongside annotations, screenshots, or other key semantic information \cite{huang_2022, rawn, stamper}. \system builds on this work in a new direction by using AI to support iteration and manage code versions.

\subsection{Support for Expressive Programming}
Creative coders work across a semantic-syntactic divide, as described by Reas~\cite{thecreatorsproject_2012}. Artists often envision and explicate their work through visual metaphors~\cite{forceville2008metaphor} or rich verbal descriptions~\cite{turner2006artful}. When using programming languages to create art, artists must then map expressive intents in the semantic space (i.e., ``undulating waves of color,'' a doodle of a design) to low-level program syntax (i.e., p5.js functions such as \texttt{lerp()} or \texttt{sin()}) \cite{thecreatorsproject_2012, ward1999drew}. However, effectively mapping semantic concepts to low-level code can be difficult for creative coders of all levels~\cite{jacobs2017supporting, victor2012}. Prior research in visual programming languages  (VPLs)~\cite{myers1990taxonomies} have looked at reducing friction by presenting some components of the code via two-dimensional visuals. For example, block-based programming languages such as Scratch~\cite{maloney2010scratch} or node-based programming languages such as TouchDesigner, Stamper~\cite{stamper}, and natto.dev~\cite{shen_2023} help users abstract important aspects such as control flow, variables, and data into visual representations. By leveraging advances in computer vision, systems such as Sketch2Code~\cite{jain2019sketch2code} directly map hand-drawn sketches to HTML code. 

Recent advancements in generative AI and text-to-code models such as Codex and Copilot have allowed users to write code using natural language prompts and autocomplete features within integrated development environments (IDEs) such as VSCode \cite{copilot_2021} and Replit \cite{replit_2022}. LLMs have accelerated the progress in bridging semantic and syntactic space by autonomously mapping natural language prompts to code output. Anecdotally, ChatGPT has already been integrated into common creative tools such as Unity and Adobe Suite. However, publically available LLMs are not necessarily optimized for a creative coding context. Mapping parameters within a high-dimensional design space to human language and feedback requires domain-specific frameworks~\cite{shimizu2020design}.

While VPLs and natural language tools provide more accessible representations of code, direct manipulation interfaces provide more accessible \textit{interactions} with code. One example of a system with a physical direct manipulation interface is Dynamic Brushes, which maps input data from a physical stylus to digital properties such as position, aesthetic style, and geometric transformations~\cite{jacobs2018extending}. Other systems involve digital interfaces for manipulation, the simplest of which involve sliders, toggles, and tuners within a block of code to allow users to tweak parameter values~\cite{2008design, victor2012}. Drawing on this body of work, \system incorporates visual node-based programming, natural language prompting with auto-complete suggestions, and direct manipulation interfaces in an attempt to bridge the semantic-syntactic divide for generative artists.

\subsection{AI Augmentation for Creative Tasks}
In creative thinking, artists draw inspiration from others' work~\cite{shneiderman2007creativity}. Research shows that novel ideas may arise from the ``genetic recombination'' of many people's work~\cite{10.1145/3491102.3501854}. In \system, we are interested in how AI, particularly LLMs, can augment human creative capabilities and inspire new forms of artistic expression as a collaborator. Text-to-Image (TTI) models such as Midjourney, Stable Diffusion \cite{rombach2021highresolution}, and DALL-E~\cite{dalle} have proliferated, but our mental models for interacting with prompt- and chat-interfaces are nascent. For example, given the stochastic nature of the models, it may take several generations of the same prompt to provide a suitable output~\cite{liuchilton}. The black-boxed nature of the TTI models does not always allow for low-level control, which may be frustrating to some artists \cite{kulkarni2023word}. One current way artists ensure somewhat reproducible and desired results is by creating a prompt template~\cite{chang2023prompt}; for example, a simple template might be ``SUBJECT in the style of STYLE''~\cite{liuchilton}. Overall, research has shown that prompt-based LLMs are useful for fast iteration and quick combination of ideas \cite{kulkarni2023word}, and early work points to potential for novel interfaces for multi-modal input, iterative inputs, and gesture-based input \cite{kulkarni2023word}. \system focuses on iterative and non-linear prompts in particular.

Auto-complete interfaces for code, such as Copilot and Ghostwriter, are becoming more powerful \cite{mozannar2022reading, barke2022grounded}. Such AI assistants help programmers in two ways:  acceleration (i.e., when the user already knows what to do) and exploration (i.e., when a user does not know what to do next) \cite{barke2022grounded}. However, it is important to consider the additional cognitive load on users to process auto-complete recommendations from the AI. When such tools were in use, the most time-consuming task for developers became ``verifying/thinking suggestions'' from the AI, which highlights the importance of a cognitively informed workflow.
\section{Formative Interviews with Experts} 
To better understand the challenges faced by creative coders and inform the design of \system, we conducted need-finding interviews with 10 expert creative coders. The study interrogated how artists currently set up their exploratory creative workflow, their approach for navigating between semantic and syntactic spaces during exploration, and how they manage iterations.

\subsection{Method}
We interviewed 10 experts in creative coding who identified as generative artists ($n=4$), software engineers ($n=2$), visual designers ($n=2$), game developers ($n=1$), and creative coding educators ($n=1$). \edit{Detailed demographic information can be found in the appendix (section~\ref{sec:demographic_info}).} Participants were recruited through word of mouth, social media posts, and snowball sampling. 7 of our participants were male, and 3 were female. All participants had 3+ years of programming experience so as not to conflate exploratory programming practices and novice programming practices~\cite{kery2017variolite}.  90\% of our participants had 5+ years of programming experience. Each semi-structured interview lasted between 45 minutes to an hour, and participants were compensated with \$25 for their time.

In the first part of the interview, we asked questions related to participants' creative processes. Example questions included ``How do you explore new creative options in your work?'', ``How do you iterate on your current work'', ``Can you tell us a time when you went back to a previous version of your code?'' In addition, we asked participants about their approach to versioning and history tracking and their experience co-creating art with AI, if any. Next, as a contextual inquiry task, we asked participants to show us a recent creative project and walk us through their virtual workspace, including code files, IDE, and output viewing interfaces. 8 out of 10 participants chose to share a creative coding project; 1 spoke about an art project in Figma, and 1 spoke about an animation project in Adobe AfterEffects. 

In the final part of the interview, we presented participants with a 20-minute open-ended creative coding task to see how they engaged in exploratory programming in real-time; the task required each participant to create a digital brush using mouse interaction based on simple starter code in p5.js. During this task, participants shared their screen via Zoom and used a think-aloud protocol to narrate their thoughts as they iterated on the starter code using Replit, an online IDE. Video recordings of the sessions were then transcribed. Two members of the research team watched all the videos and included relevant screenshots at appropriate points in the transcript documents. The same two members then read through all of the interview transcripts and conducted inductive coding~\cite{strauss1990basics} on ATLAS.ti~\cite{atlasti} and identified key insights related to iteration, creativity, and exploratory programming. Through multiple rounds of discussions, we clustered the insights into high-level themes via affinity clustering and thematic analysis~\cite{braun2006using}.

\subsection{Findings and Design Considerations}

\subsubsection{Parameter- and accident-driven exploratory programming} 
Many of the participants described their exploratory process as parameter-driven, in which their exploration is based on tweaking a set of variables, which U9 referred to as ``magic constants.'' U4 organizes their code by placing all of their parameters at the top of the file: \textit{``I don't really have a great standing process for tracking versions, and I kind of just make a mental note, or just write down  parameters that lead to something that I like\ldots I like to chunk and get all the things that I would be modifying in one place and get all the things that are fairly static out of the way."} When asked how they knew that they were iterating on parameters in the right direction, many participants pointed to previous experience. They explained that experts typically drew from previously used parameters from prior work. One commonly used pattern that experts used in the open-ended task involved constraining circular movement using the parameter range of $-\pi/4$ to $\pi/4$ radians. Therefore, CSTs should allow \textbf{artists to maintain fine-grained control over the parameters they are manipulating (Design Consideration 1).}

Other times, iterations on parameters came about through a process of trial-and-error, which some participants termed ``accident-driven development.'' U3 describes a time when she uncovered a new idea in a p5.js sketch: \textit{``I messed up what I was connecting, and then I ended up with two objects connected in the middle\ldots so I think the little accidents are the inspiration.''} During such discovery, surfacing parameters while running and testing code was another point of conversation; this was sometimes done through console statements or small helper functions. U7 described writing ``debug'' functions which programmed hotkey buttons to show him the range of possible outputs while he was running a 3D game simulation, especially in situations where controlled randomness was required. U5 described the cycle of iterating on a particular parameter, running the code, and analyzing the result as a feedback loop: \textit{``If I had like a little slider to make [this particular parameter] more or less like I was just tuning constantly\ldots If there was a way to make that feedback loop quicker, that'd be perfect.''} Thus, CSTs should \textbf{allow participants to read and manipulate parameter values (D2) easily} and \textbf{readily see the output to support rapid creative exploration (D3).}

\subsubsection{Version control practices}
Because creative exploration is often brought about by unintended changes, participants described mental friction between the exploratory process and the intentional process of committing and summarizing changes via version control systems like Git. In both parameter- and accident- driven development, tightly bound cycles of iteration, execution, and analysis were observed:  \textit{``So I do a lot of it in my head or like taking notes and stuff like that, and so I’ll run it [and] see what happens''} (U7). Most participants chose not to use a version control system when creating personal projects despite having familiarity with Git. U10, one of the few who did use Git for creative work, described the friction using Git commits and branches in this way: \textit{``It just required too much planning ahead when this process [creative exploration] is not about planning ahead; it's about seeing where you go.''}

Instead, many participants described their informal versioning practices (similar to \cite{kery2017variolite}). As a workaround, participants would duplicate files or code snippets and comment them out when not in use. Alternatively, three other participants described a process of tracking changes in working memory or by jotting notes down, saying some variation of ``I just remember what I did previously'': ``\textit{I do a lot of like\ldots open up notepad and write down what was good [while running], and then, just remember that so it's definitely more of a conceptual than a version control thing}'' (U7). Therefore CSTs should provide \textbf{versioning and tracking as an integrated experience with the rapid exploration process (D4).}

In addition to tracking code configurations that produced salient outputs, participants also used informal versioning practices to keep track of the visual outputs of the code.  Participants would track changes by running and storing the outputs via screenshots, and video captures (e.g., ``versionFinal.jpg'', ``versionFinalFinal.jpg''). While these were arguably lower fidelity than Git commits that included code \cite{sterman2022towards}, participants found this helpful to capture the semantic constructs that they were experimenting with at the time. As U3 described: \textit{``So the way I iterated was whenever I had something distinct enough that I wanted to save, I would just duplicate the p5 file and name it, then move on\ldots So, from one generation to the other, I probably just saved it duplicated and then started working on top of that. It's very jank version control.''} However, the outputs were often detached from code or any other useful history-related metadata. In this way, CSTs should provide ways to effortlessly \textbf{track the emergent visual outputs during exploration along with the underlying code that produced the output (D5).}

\subsubsection{Interest in AI Augmented Creative Exploration}
The interviews were conducted in the summer of 2022. At that time, ChatGPT was not yet publicly available, but most artists had heard of and experimented with TTI models, code generation with Copilot, and other LLMs. Most participants were unsure how they felt about using AI in their personal projects. U3 stated that she ``used to be very much against it'' and would only consider using it after the bulk of the creative work was completed. U10 similarly felt AI would be good for ``post-processing.'' He thought about AI as a way to add \textit{``filters''} on top of his work:  \textit{``I don't want to use [AI] for the core of my thing, but I could see using it for enhancing whatever I just did. So like if I created some pattern, I could [ask AI to] give it a great bloom effect or add a bunch of film grain.''}

Participants did feel comfortable and excited about using AI to generate expressive prompts for inspiration, as it could help them explore new creative directions and overcome creative blocks. When asked if they could imagine iterating on creative code with a computer, they emphasized the importance of the human leading the creative process. U9 compared the utility of AI to a randomize feature that provides suggestions for new parameters for him to try. U10 suggested that the AI could be trained on his personal history: \textit{``Maybe it could learn from myself. All the other tools are learning from the world at large and existing art. I wonder if it could learn from the way that I do things. I wonder if it could be throwing up suggestions or provocations\ldots side by side [while I code].'' } 

In summary, AI-augmented CSTs should \textbf{offer different types of AI support for different parts of the creative process (beginning vs. iteration vs. end) and allow the artist to maintain creative control at all times (D6)} .
\section{User Experience}
 \setlength{\fboxrule}{1pt}

\begin{figure}[htb]
  \centering
  \includegraphics[width=\columnwidth]{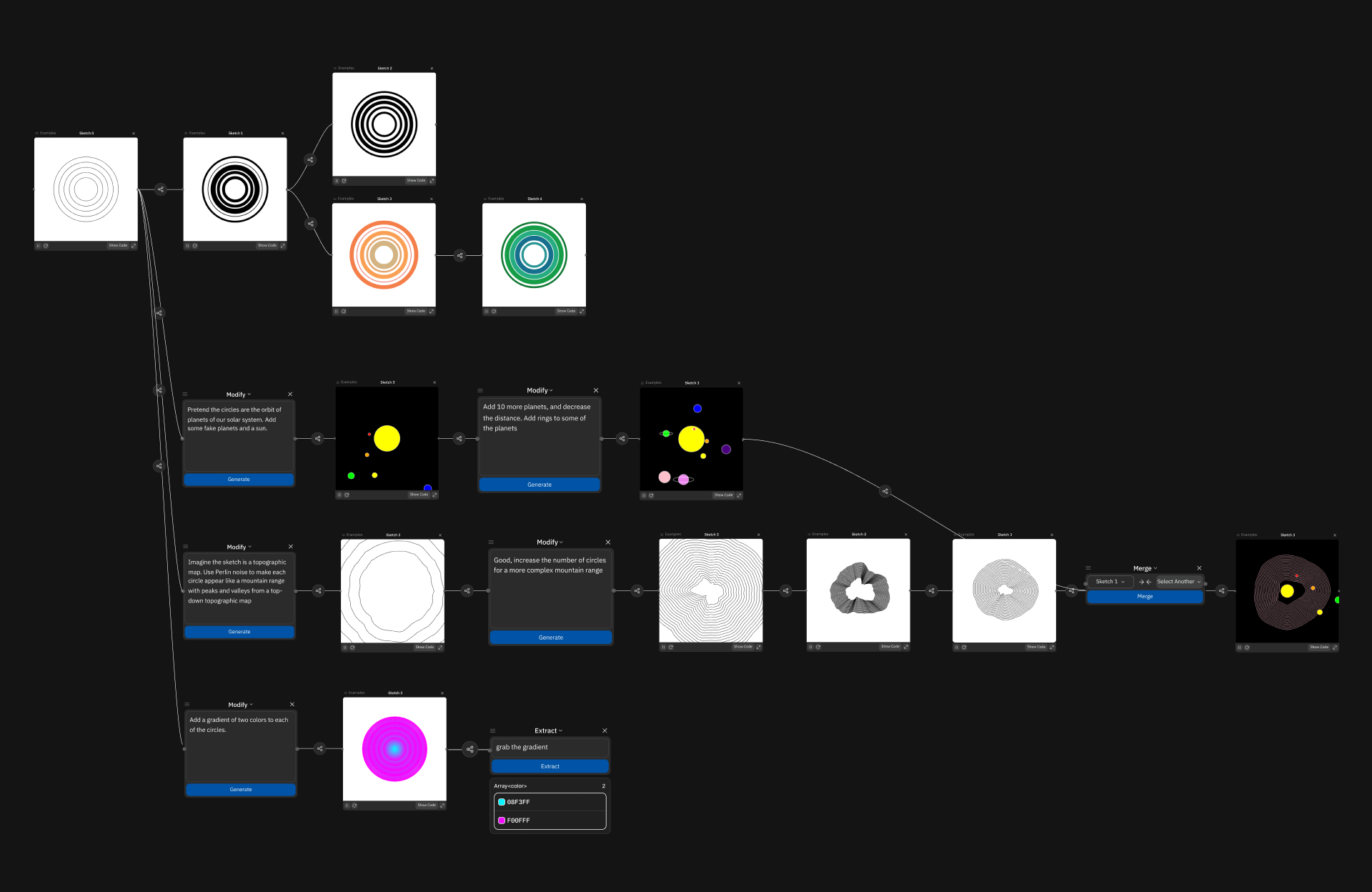}
  \caption{Overview of Spellburst's User Interface}
  \label{fig:interface-overview}
\end{figure}

\begin{figure}[htb]
  \centering
  \includegraphics[width=.9\columnwidth]{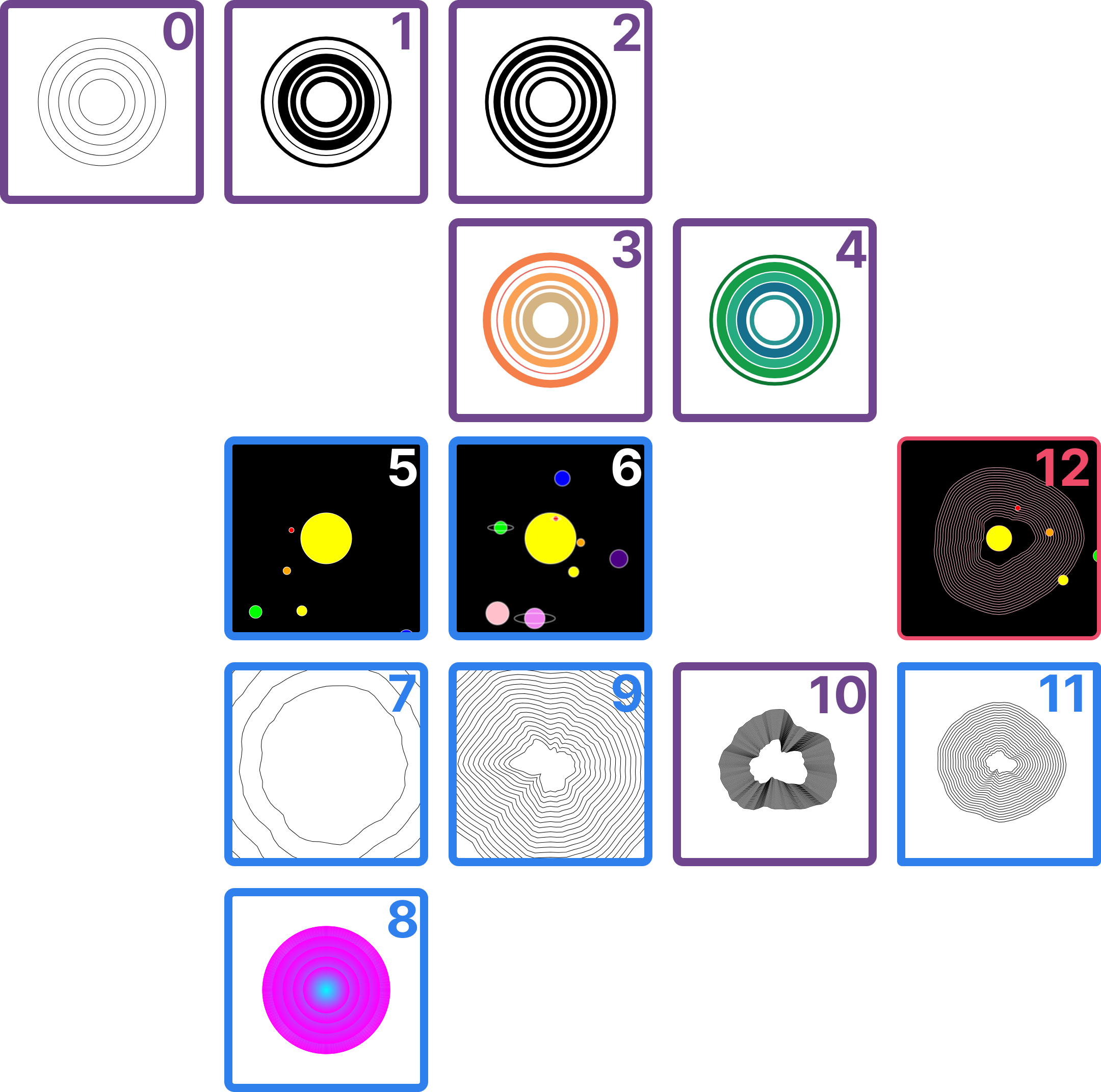}
  \caption{Sketches created by Ash during an exploratory creative coding session, color-coded by node type.}
  \label{fig:ux_sample}
\end{figure}

\system is designed to facilitate rapid exploration and visualization of a large tree of possibilities while simultaneously preserving exploration history (D4 and D5). As shown in Figure~\ref{fig:interface-overview}, \system's user interface consists of an infinite zoomable canvas containing one or more \textit{node-based} layouts. The node layouts follow a hierarchical tree structure, and each node is either a \fcolorbox{sketch}{white}{sketch} or an ---\fcolorbox{black}{white}{operator}\textrightarrow\xspace. \fcolorbox{sketch}{white}{Sketch} nodes represent p5.js~\cite{p5js} programs that are compiled and rendered to a canvas. The root node indicates a starting sketch for a creative path. ---\fcolorbox{black}{white}{Operator}\textrightarrow\xspace nodes are \textit{links} that indicate the following exploratory programming operations: modification, merging, duplication, diffing, and extraction. Some of these operator nodes allow natural language inputs; all of them output new sketches based on source sketches and prompts.  

To demonstrate our system's features and user experience, we describe how Ash, a creative coder, creates and iterates upon a generative art project using \system.

\subsection{Set-up}

\begin{figure*}[htb]
  \centering
  \includegraphics[width=\textwidth]{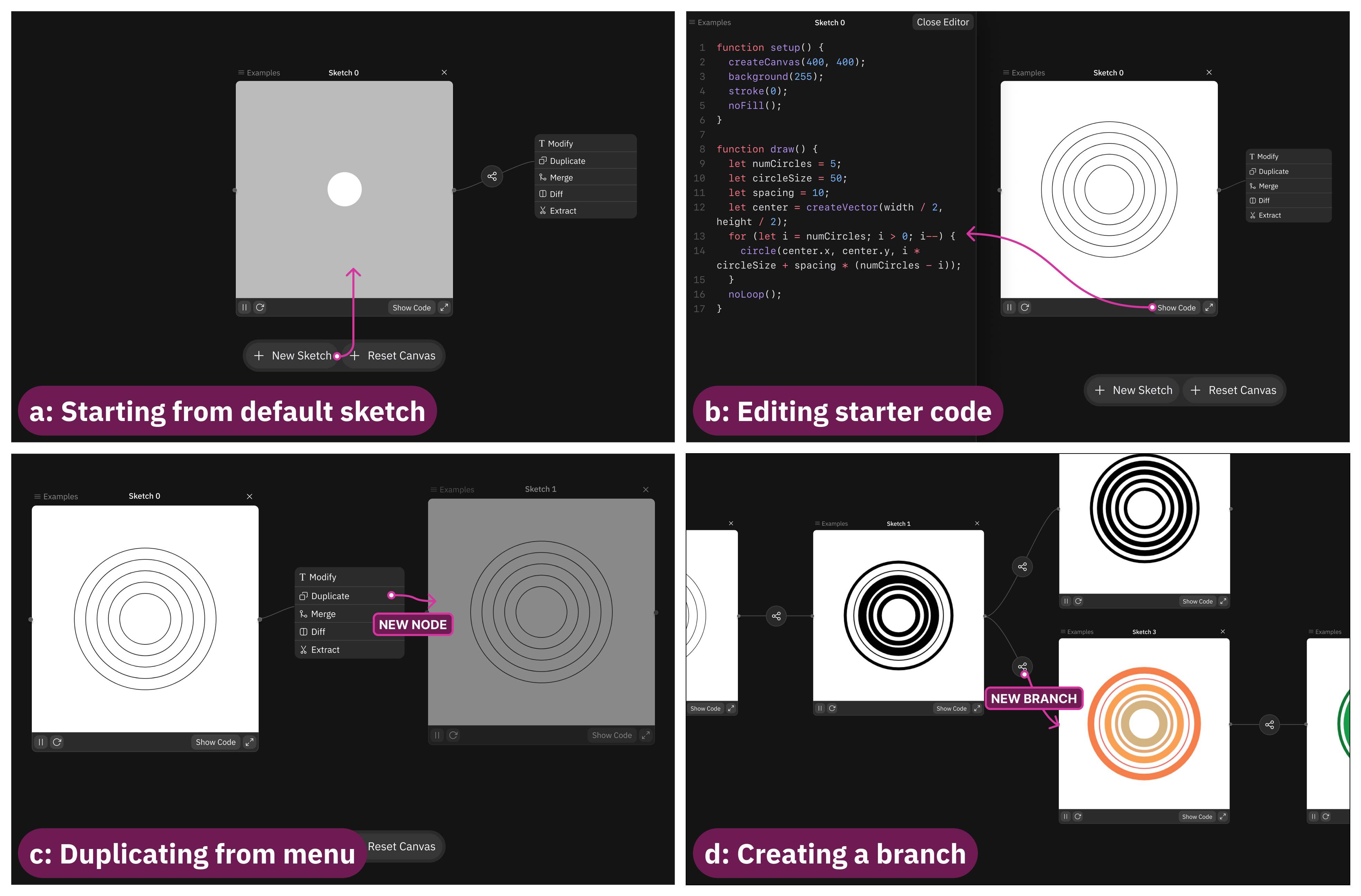}
  \caption{Setup and basic high-level controls of \system}
  \label{fig:ux1}
\end{figure*}

Ash has a conceptual idea about \textit{concentric circles} to inspire her exploration. She opens \system on her web browser, which presents a blank canvas with a floating add node `+' button (Figure~\ref{fig:ux1} a). Ash clicks on the button to create a root node, which is a \fcolorbox{sketch}{white}{sketch}. Next, she clicks on the `show code' button on the root node, which brings up the code editor to the left of the canvas (Figure~\ref{fig:ux1} b). Here, she can author an initial sketch from scratch or copy code for concentric circles from sources on the web (such as the p5.js~\cite{p5js} gallery). Ash pastes the p5.js code snippet to render four concentric circles from the web, which displays the output on the root node (Figure~\ref{fig:ux_sample} Sketch 0). \system also has an `Examples' tab with a set of starter code for basic shapes.

\subsection{Focused Exploration}
Starting with the initial root \fcolorbox{sketch}{white}{sketch} node, Ash clicks on the button on the right edge of the node and selects the ---\fcolorbox{black}{white}{duplicate}\textrightarrow\xspace operator in the popup menu (Figure~\ref{fig:ux1} c). This creates a copy of the root node and is visualized as a child \fcolorbox{sketch}{white}{sketch} node to the root node. \edit{Compared to platforms like Fork It {\cite{10.1145/3411764.3445527}}, Spellburst is designed to encourage iteration and ``accident-driven development'' rather than just executing nodes in parallel. Thus, all {\fcolorbox{sketch}{white}{sketch}} nodes are connected to an operations menu, where selecting an option is required to continue adding nodes to the branch rather than being purely open-ended.} 

In the newly created copy, Ash modifies the sketch's code in the code-editor to algorithmically randomize the line thickness. By repeatedly making copies of the previous node, Ash explores different approaches to varying line thickness (Figure~\ref{fig:ux_sample} Sketches 1 and 2). At this point, Ash decides to explore color variations.

However, she wishes to explore colors separately from line thickness. Selecting one of the previous sketches as the base, she initiates a $\xrightarrow{\text{branch}}$ operation by selecting from the button on the edge between the original sketch and its duplicate node (Figure~\ref{fig:ux1} d). \edit{Note that unlike platforms like natto.dev~\cite{shen_2023}, our edges aren't freeform and arbitrary. They are generated from duplicating and iterating on existing sketches and prompts rather than being derived from code dependencies.} In this new branch, Ash explores variations of color by directly editing the attributes and functions in the code editor, which produces Sketches 3 and 4 in Figure~\ref{fig:ux_sample}. As we will see later, Ash can also use natural language prompts to execute these changes, but based on our interviews, artists prefer greater control over familiar attributes (D1 and D6).

\begin{figure*}[htb]
  \centering
  \includegraphics[width=\textwidth]{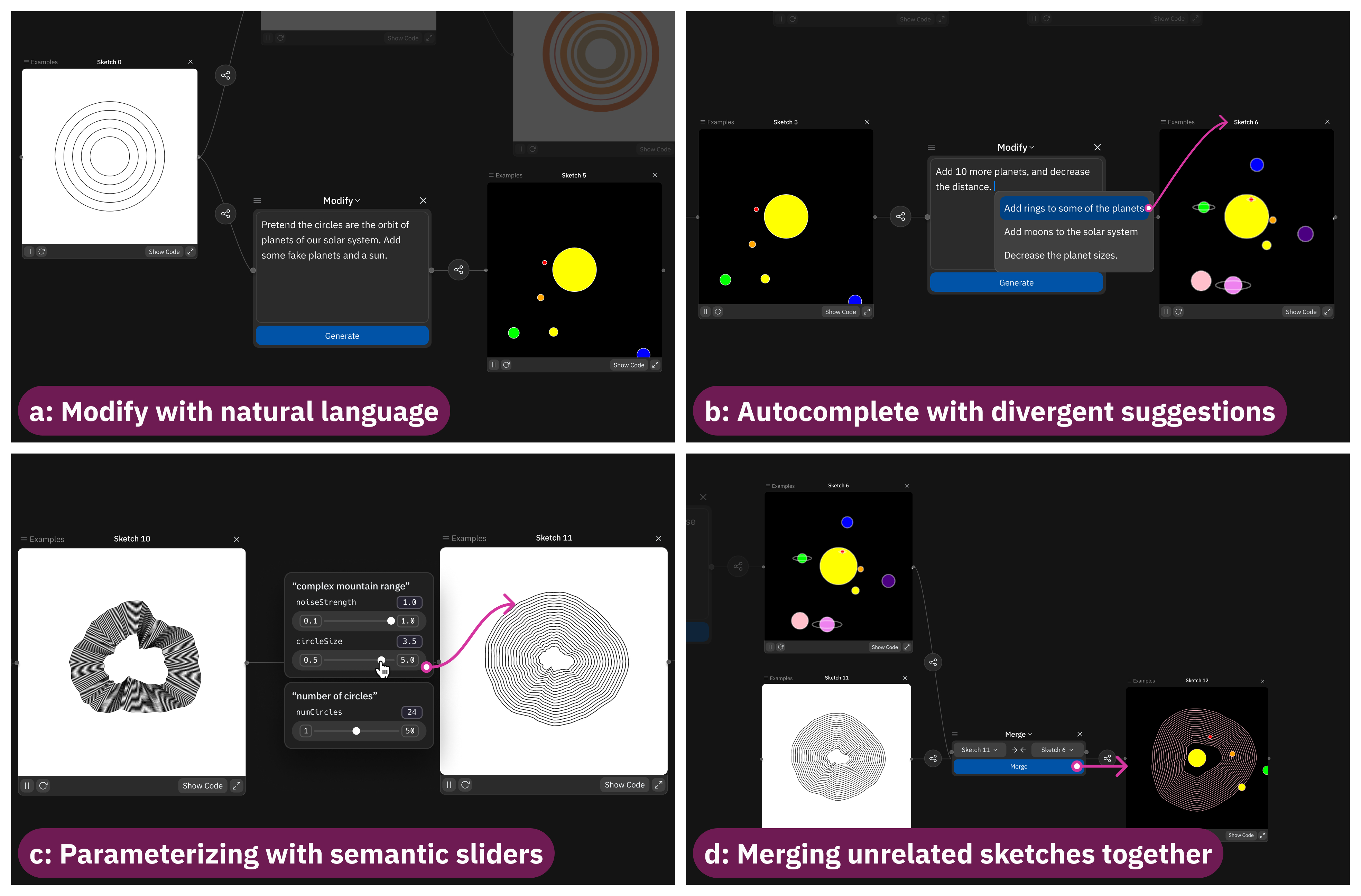}
  \caption{Modification, merging, and semantic sliders powered by LLMs}
  \label{fig:ux2}
\end{figure*}

\subsection{Exploring Large Creative Jumps}
At this point, Ash imagines a new idea of circles representing the planetary orbits of our solar system. She decides to use \system's natural language prompt  to explore this new direction. Clicking on the initial \fcolorbox{sketch}{white}{sketch} node, Ash initiates a ---\fcolorbox{modify}{white}{modify}\textrightarrow\xspace operator, which brings up a prompting interface along the link (Figure~\ref{fig:ux2} a). In the text box, Ash enters the prompt \textit{``Pretend the circles are the orbit of planets of our solar system. Add some fake planets and a sun.''} Using the previous sketch and the prompt as inputs, \system generates a new \fcolorbox{sketch}{white}{sketch} node with an animated output showing several concentric circles and elliptical objects of different sizes animating along the orbits (Figure~\ref{fig:ux_sample} Sketch 5). Such a large creative jump would have been time-consuming to manually program. 

Ash inspects the generated output and finds it interesting, so she decides to continue with this new direction. She creates a new ---\fcolorbox{modify}{white}{modify}\textrightarrow\xspace operator and begins to write a new prompt: \textit{``Add 10 more planets, and decrease the distance.''} To help Ash with her exploration, \system provides inspiration through an auto-complete drop-down (Figure~\ref{fig:ux2} b), which suggests three additional ideas: \textit{``Add rings to some of the planets,'' ``Add moons to the solar system,'' and ``Decrease the planet sizes.''} She hadn't thought about adding rings, but she likes the idea and selects this auto-complete option resulting in a new \fcolorbox{sketch}{white}{sketch} node (Figure~\ref{fig:ux_sample} Sketch 6).

Later, Ash returns to the original sketch and explores two entirely different concepts: one related to Perlin noise and topographic maps (Figure~\ref{fig:ux_sample} Sketches 7,9,10,11) and another related to color gradients (Figure~\ref{fig:ux_sample} Sketch 8). These are depicted as two additional branches from the initial sketch node. Ash uses the prompts \textit{``Imagine the sketch is a topographic map. Use Perlin noise to make each circle appear like a mountain range with peaks and valleys from a top-down topographic map''} to produce Sketch 7 and \textit{``Good, increase the number of circles for a more complex mountain range''} for Sketch 9. 

\subsubsection{Iterating on generated outputs through code editing}
By looking at the output for the topographic maps, Ash is not quite satisfied with Sketch 9. In this instance, prompt-based refinement can be tedious. Therefore, Ash decides to iterate directly on the code on a more granular level. As before, she clicks on the `show code' button to open up a code editor view of Sketch 9. \system automatically provides code comments for the prompt-generated code. She clicks on the `maximize' button to focus the interface  and zoom in on just that node ( similar to Figure~\ref{fig:teaser} c). She manually updates the values for the \texttt{spacing} parameter to create a denser arrangement, which instantly updates the output view. Once Ash is satisfied with this version, she zooms back out to the canvas, and adds a ---\fcolorbox{black}{white}{duplicate}\textrightarrow\xspace operator to make a copy of Sketch 9 so she can manually edit the code; she creates Sketch 10 this way.

\subsubsection{Iterating on prompts using Semantic and Global Variable Sliders}
To provide fine control over the prompt phases or propositions, \system has auto-generated various sliders for global variables present in the code: \texttt{circleSize}, \texttt{numCircles}, and \texttt{noiseStrength}. Further, she notices that \system has automatically mapped relevant words and phrases from the original modification prompt to these variables, allowing her to adjust the relative influence of words in the prompt and see changes to the sketch in real-time (Figure~\ref{fig:ux2} c). Ash plays around with both of the sliders, producing Sketch 11. \edit{These dynamic prompt-driven sliders, in addition to our non-deterministic autocomplete, allow Spellburst branches to easily enable quick exploration of extremely divergent creative paths. Such affordances highlight Spellburst's capacities for iteration and ``accident-driven development'' unlike platforms like Fork It.}

\subsection{Merging Nodes and Extracting Properties}
As Ash iterates on her project, she realizes that she wants to combine visual aspects from two previous sketches along two different branches. However, manually merging code from different sketches can be time-consuming. Instead, Ash uses \system's ---\fcolorbox{merge}{white}{merge}\textrightarrow\xspace feature to select Sketch 6 and Sketch 11 for merging (Figure~\ref{fig:ux2} d). She does not provide any natural language input; moments later, Sketch 12 appears on screen. When Ash inspects the code for Sketch 12 in the code editor, she finds the following code comment, which is a natural language description of how the two sketches were merged: ``Combine the planet system from [Code Snippet 1] with the mountain range background from [Code Snippet 2] which produces Sketch 12. The resulting sketch shows a planet system with a mountain range background that updates in real-time and is generated using Perlin noise." 

In addition to merging sketches, Ash can use the prompt interface to extract specific sketch attributes as a new sketch (different from duplicating the entire sketch). For instance, Ash uses the ---\fcolorbox{black}{white}{extraction}\textrightarrow\xspace opreator to isolate the color gradient from Sketch 8. She's not sure where she plans to use it, but she wants to keep it readily accessible. In a different step in the exploration workflow, Ash can take a step back from the work as is and look at differences between the wide variety of sketches currently on the canvas. She takes the latest output from Sketch 11 and creates a new ---\fcolorbox{black}{white}{diff}\textrightarrow\xspace operator to see a natural language summary of how Sketch 11 has changed from Sketch 10, as well as the differences between the sketches at a code level. She can then use this summary as input for future modifications.

Ash notices that \system will occasionally generate code that does not run or code that conflicts with her expectations. Whenever this happens, Ash has two options: (1) step into the code editor to debug the code manually or (2) delete the node and regenerate the result. She does this various times throughout her session, pruning and growing her branches with ease.  \edit{While the zoomable canvas gives her a ``birds-eye view'' similar to other node/canvas UIs, our structured auto-layout removes the need to ``clean up'' the canvas and makes it immediately clear how a sketch has progressed over time because it is a direct visualization of the version history tree. We also chose to focus on auto-layout so that artists would be focused on iteration rather than organizing their canvas, which was based on cognitive load concerns in our formative study.}

Throughout her work with \system, Ash benefits from the seamless integration of syntax and semantics, the intuitive Node View, and the powerful natural language interface. The system's features enable her to focus on her artistic intentions and iterate on her generative art project with greater ease and control. By the end of this session, Ash has only begun her creative journey. She has successfully explored various corners of the design space, as sampled in Figure~\ref{fig:ux_sample}. 
\section{\system}
\label{sec:architecture}
The core of \system's implementation focuses on chaining together API calls to OpenAI~\cite{chatgpt} to generate p5.js sketches. This involves dynamically composing the prompt context based on manually written code, AI-generated code, previous prompts, and existing operator node data. Here we describe the prompt engineering approach for the key functionalities of \system and provide details about our system implementation. The full set of prompts used in \system are provided in the appendix (section~\ref{sec:base_prompts}). 

\subsection{Prompt Design}
To design the prompts, we iteratively explored a combination of prompting strategies based on a trial-and-error approach, \edit{using visual output as our guide. Unlike most LLM-powered coding assistants that mostly automate coding for one-off tasks, our prompts are intended to be ``bridges'' between p5.js source code and their visual outputs.}

\subsubsection{Code Generation}

\edit{When first experimenting with prompting GPT-3.5 to output \texttt{p5.js} code, we used simple one-sentence questions (e.g., ``Draw a \texttt{p5.js} code snippet that shows\ldots''). However, we found that the model provided either code that did not work or output that did not align with our expectations. To address this shortcoming, we employed \textit{few-shot prompting.} Few-shot prompting is derived from the technique of few-shot learning, where a model is trained on several related tasks in order to generalize well on new tasks with just a few examples {\cite{10.5555/3495724.3495883}}. This technique incorporates several examples of desired input and output examples within the prompt to steer the model's output in the right direction. Our examples consisted of both simple code snippets (e.g., a circle and a square on a screen) to more complex code snippets that allowed our model to decode abstract concepts (e.g., balls that bounce off the walls of the edges of the canvas). Below is a snippet of one of the examples we provided within our few-shot prompts:} \\

\begin{lstlisting}[backgroundcolor = \color{beige}, mathescape=true]
$\textbf{Prompt: add a bunch more balls and make them bounce off the bounds}$

$\textbf{Output:}$
let numCircles = 20;
// Create an empty array to store the circles
let circles = [];
// Set up the canvas and create the circles
function setup() {
  createCanvas(700, 410); 
  ...
\end{lstlisting}

\edit{After using few-shot prompting, we wanted to integrate the model's outputs within the \system system. To create usable output, we forced the model to provide its answer using a specific \textit{template} and \textit{syntax}. At a high-level, templates are an effective prompting strategy, as they filter LLM output to eliminate choices that would have been unhelpful to the user~\cite{white2023prompt}. For instance, our prompts contained start and end tokens as well as extra instructions to make sure the code compiled and rendered correctly (e.g., limiting calls to the \texttt{draw()} function to ensure animation outputs render correctly). Getting LLMs to generate structured data consistently is an open research effort, with projects like jsonformer \cite{sengottuvelu_2023} or grammar-based sampling \cite{ggerganov_2023} allowing users to specify output schemas. However, we used the following restrictions in our prompt to guide the output:} \\\\

\begin{lstlisting}[backgroundcolor = \color{beige}, mathescape=true, language={}]
$\textbf{Restrictions:}$
- Only respond with code in your output as a raw string. 
- Be as efficient as possible with your implementations. When producing computationally intensive sketches, try to use optimization methods so they run more quickly. 
- If you are ever asked to apply an animation, remember to always remove any calls of the noLoop function to make sure it actually animates.
- Comment your code with useful comments.
- Remember to be as efficient as possible with your implementations. When producing computationally intensive sketches, try to use optimization methods so they run more quickly. 
\end{lstlisting}

\edit{Finally, we ran into specific problems that were unique to our use case of creative coding. While the use of natural language queries was helpful in making large jumps within the semantic space, the lack of the model's explanation made it difficult to make more granular syntactic edits. To make the model output more explainable to users seeking finer control, we experimented with prompting the model to write code comments. We embedded these code comments within our few-shot prompts, outlining the purpose of every couple of lines. In this way, the code blocks served as chain-of-thought prompts, encouraging the model to think ``step-by-step'' to decompose hard tasks into smaller steps {\cite{NEURIPS2020_1457c0d6}}.  
}

\subsubsection{Prompt Auto-complete}
Prompt authoring tools are currently underpowered and often involve simply typing into inputs without any form of auto-complete or facilitation for future direction. Unlike auto-complete for structured data like a programming language -- where the system can easily match the input with a pre-existing set of options -- natural language input can have multiple interpretations. Thus, it can be difficult to determine what the ``correct'' completion is, especially for creative text-generation tasks. Because of this ambiguity, we focused the \system auto-complete on providing non-deterministic options to suggest interesting directions to take a prompt. Supporting exploration, rather than just ``task execution'' or search, is also important. By presenting a range of possible suggestions, auto-complete can encourage users to think more deeply about their search and discover new information that they may not have otherwise considered. The goal here is breaking past creative barriers and novelty.

With these design decisions in mind, we implemented prompt auto-complete by querying ChatGPT, \edit{again by using few-shot prompting to make the API simulate being an auto-complete engine}. For context, we fed in the connected sketch to extract relevant keywords from (1) the current prompt being typed and (2) 5 examples of input-output prompt completion pairs.

\subsubsection{Semantic Parameter Adjustment}
Semantic parameters allow users to adjust sketches at the concept and prompt level, rather than needing to search for exactly which variables to manipulate in the source code. Our primary implementation method involves \edit{(1) extracting out key phrases from the input prompt, (2) generating a "semantic map" between these phrases and related global variable names, and (3) generating the desired code from the initial prompt and the semantic map with another API call.} We found success by telling the LLM up-front to first name the global variables it would declare in relation to keywords in the prompt. This method mimics the data you might get from an attention ``heatmap'' between the input prompt and the output code, which is currently not possible to obtain due to the closed-source nature of OpenAI's LLMs.

\subsubsection{Merging Operations}
Our implementation of merging is different from traditional merging algorithms. Existing methods, such as the three-way merge used in Git, try to combine code contents line-by-line without producing syntax errors. On the other hand, \system's merge functionality is LLM-powered, meaning that it tries to combine code contents in a semantically meaningful way. For example, it may take the physics behavior from a bouncing ball simulation and ``merge'' the color palette from an unrelated abstract art sketch. We call this process \textit{semantic merging}. 

\edit{To implement semantic merging, we used few-shot prompting to provide examples of expected outputs, a template and syntax to integrate the model's response within \system, and code comments to make the code explainable. One area we intend to explore more involves what a few-shot prompt example could look like. In some tests, we simply juxtaposed two \texttt{p5.js} sketches next to each other. For other tests, our merges were larger semantic jumps that would be hard to achieve by merely directly combining the two code snippets together. An example of one of our merging prompts can be found in \ref{sec:base_prompts}.}

\subsubsection{Developing a Taxonomy for Creative Coding Transformations}
Finally, to align the natural language queries from users and the expected output from ChatGPT, we developed a taxonomy that classified different aspects of creative coding transformations. This taxonomy was developed by examining a variety of artistic taxonomies and adapting them to the context of creative coding. We organized the taxonomy into three main categories of transformation:
\begin{enumerate}
    \item \emph{Apply Transformation to Objects / Primitives / Marks}: This category deals with the individual elements or building blocks used in creative coding projects. These include properties such as color, shape, form, texture, thickness, waviness, curviness, and randomness*.
    \item \emph{Apply Transformation in Relation to the Plane/Canvas}: This category addresses how the elements in a creative coding project are transformed concerning the overall composition or canvas. Properties in this category include size, direction/orientation*, alignment*, white space*, movement, noisiness, symmetry*, scale/proportion*, hierarchy*, and randomness*.
    \item \emph{Apply Transformation to the Relationship Between Objects}: This category focuses on the relationships between elements in a creative coding project and how they can be transformed. Properties in this category include nesting, direction/orientation*, repetition/pattern, proximity/spacing, alignment*, white space*, symmetry, contrast/emphasis, variety, balance, scale/proportion*, hierarchy*, unity, depths/layers, and randomness*.
\end{enumerate}

Asterisks indicate properties that appear in two categories, illustrating the interconnectedness of these properties in creative coding transformations. 

\edit{With this taxonomy, we searched and reviewed over 40 examples of creative coding work and labeled them according to this categorization. Subsequently, we wrote a code snippet corresponding to a change a participant could suggest in alignment with the taxonomy. Lastly, we crowd-sourced natural language queries that could be used to describe the transformation of the code from its original state to the state we coded. A curated number of crowd-sourced responses were then used as seed data to generate the few-shot examples for the ChatGPT API calls. Details on the crowd-sourced survey can be found in {\ref{sec:image_transformation}}.}

\subsection{Implementation Details}

\subsubsection{User Interface:} The frontend of \system is built using a combination of \texttt{React}~\cite{react}, \texttt{Reactflow}~\cite{reactflow}, \texttt{Jotai}~\cite{jotai}, and \texttt{d3.js}~\cite{d3js}. \edit{\textit{React} is the foundation for our user interfaces, while \texttt{Jotai} allows us to store and manage the state of the graph's nodes and edges. For visualization, we used \texttt{Reactflow} to build the node-based interface for exploring variations through branching and merging, while \texttt{d3.js} is used to manage the hierarchical tree visualization and auto-layout of the graph.}

Because the canvas involves constantly running multiple sketches at the same time -- each of which with the ability to individually pause, play, and reset -- we run each sketch in a separate \texttt{iframe}. To make the editing experience fast and responsive, we reload a sketch's \texttt{iframe} on every keystroke inside of the code editor. The most important quality of this automatic reloading is making sure the system is reactive and users can instantly respond to any visual feedback, which is important for discovering unexpected properties or behavior in sketches.

\subsubsection{Backend and Language Model API:}
The \system backend is built using \texttt{Node.js}~\cite{node} and \texttt{Express.js}~\cite{express} and uses OpenAI's ChatGPT~\cite{chatgpt} API for all language model functions, specifically the~\texttt{gpt-3.5-turbo} model. The backend has several routes for modifying, merging, extracting, and auto-completing code, each dynamically generated based on the input. 

\subsubsection{\system Graph:}
Each node and edge in the graph is represented as a JSON object. The properties of the respective JSON schemas help define the position, source code, and connections to other nodes on the graph. An example of the JSON objects for both components can be found in~\ref{sec:json_schemas}. All operations on the graph are immutable by default, meaning that all modifications and adjustments to source code result in the creation of new nodes rather than overwriting the input. This immutable structure allows users to easily view the causal dependencies between operations and how sketches progress over time.

Whenever an operator node is re-run but is already connected to existing sketches, the connected sketches will automatically update with the new code. Regenerated code only affects immediately connected sketches/edges one layer deep, so ``out of date'' code needs to be re-run one layer at a time. When a node is deleted from a graph, its descendants will remain in the graph. However, the descendants will be reattached to the deleted node's parent node. Although this method does not preserve the original causal structure of the graph, it facilitates visualization of the overall progression of how the initial sketch diverged over time.
\section{Expert Evaluation}
In our evaluation, we aimed to determine whether artists are able to engage in exploratory creative coding using \system. Concretely, we gathered feedback on \system's support for (1) bridging semantic and syntactic spaces while creative coding through natural language prompting, (2) control over creative coding across large and small code increments, and (3) the usability and cognitive load while using \system.

\subsection{Method}
We conducted a user study via Zoom with 11 expert creative coders. We recruited participants by contacting generative artists in our professional network and through social media messaging. One participant was a repeat participant from the formative study. Each session lasted approximately 75 minutes, and participants received \$30 for the time. \edit{Participants' demographic information can be found in the appendix (section~\ref{sec:demographic_info}).}

In each session, participants were given a walkthrough tutorial of \system's key features and asked to complete a practice creative coding task. Once participants indicated familiarity with \system, they proceeded to work on a set of creative coding tasks that required performing small changes to an initial starter code using prompts or direct code editing (e.g., tweaking the color of the circles in the sketch, four tasks total), executing large creative shifts using prompts (e.g., proposing an iteration that is a significant departure from the current code, four tasks total), and open-ended creative exploration using \system's feature for 15 minutes. \edit{The juxtaposition of small and big changes maps to syntactic and semantic changes, respectively. By designing our two tasks like this, we wanted to formalize the system's ability to allow people to move between abstraction levels.} Participants completed distinct tasks in each category using \system and their typical IDE alongside ChatGPT~\cite{chatgpt}. We ran this as a within-subject comparison study, and the order of tool use was randomized. At the end of the study, each participant completed the NASA-TLX cognitive load questionnaire, a usability questionnaire, and finally participated in a 15-minute long open-ended discussion. 

\subsection{Findings}
\begin{figure*}[htb]
  \centering
  \includegraphics[page=1, width=0.8\textwidth]{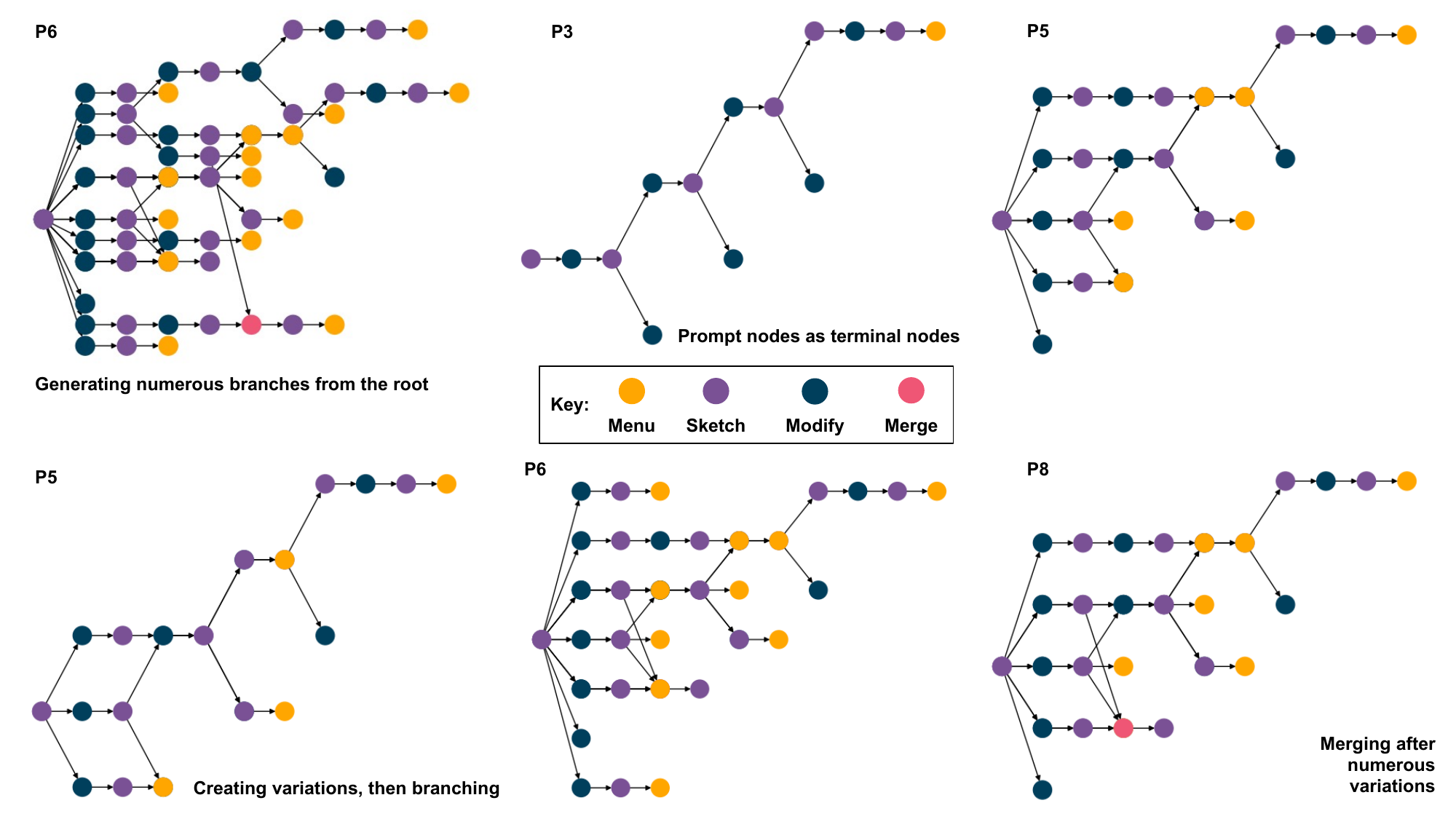}
  \caption{Abstract node visualizations of the canvas. The collated examples showcase the key affordances and behaviors associated with \system.}
  \label{fig:recreated}
\end{figure*}

Using the snapshot of the canvas at the end of each user's run -- encompassing the nodes, edges, and their types -- we created a visualization of artists' exploration. The resulting visualization in Figure~\ref{fig:recreated} shows distinct patterns of creative explorations for several participants in \system, and appendix \edit{\ref{sec:example_output} shows screenshots of Spellburst from two participants}. Across tasks and open-ended exploration, participants engaged in rapid iteration, which ranged from fine-grained editing to large semantic changes. Expansive patterns of divergent thinking were encouraged through \system's affordances with multiple creative pathways emerging from the root node. This can be especially seen for P6 in the upper left portion of Figure~\ref{fig:recreated} which involved 10 different modifications from the root node, and P8, who merged after creating numerous variations. \system can support different variations of depth and breadth in exploration, allowing users to easily explore a new area if they find that one line of iteration is no longer relevant.

Further, our analysis shows three types of creative exploration:

\begin{enumerate}
    \item \textit{Branching and Variations:} For some, the optimal way to use \system was to create numerous branches off of the root node and then perform numerous modifications and merges from each branch. On the other hand, other participants found it beneficial to make either one or a few branches from the root node, perform a small number of modifications, and then create even more branches from those.
    \item \textit{Modify Nodes as Terminal Nodes:} While the majority of the terminal nodes are menu nodes -- indicating sketches -- a good number are also modifying nodes, with prompts to create variations that were ultimately unexplored.
    \item \textit{Merging after Several Modifications:} Merging often occurs after several modifications of a branch of changes. As a result, users often prefer to explore the possibilities within branches before merging two together, as opposed to merging together after smaller changes.
\end{enumerate}

\subsubsection{Prompt Based Editing}
\begin{figure*}[htb]
  \centering
  \includegraphics[width=\textwidth]{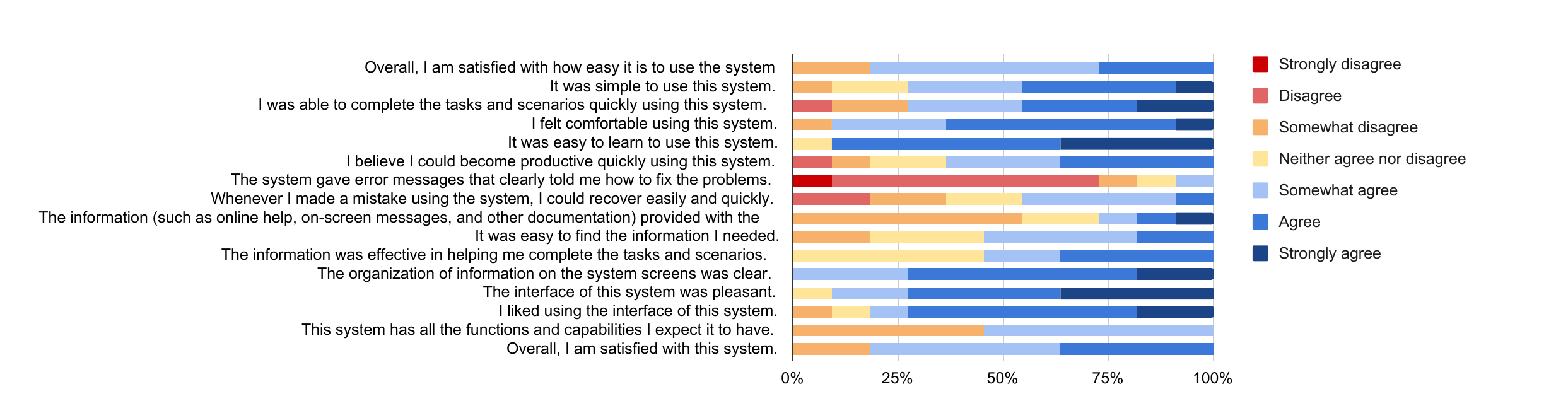}
  \caption{Participant Responses to Post-Study System Usability Questionnaire (PSSUQ)}
  \label{fig:pssuq}
\end{figure*}

Across all sessions, participants responded positively to using prompts to initiate larger creative shifts. \edit{Although many expressed that it was their first time using LLM prompts in creative coding work, 73\% agreed or strongly agreed that they liked using \system's interface and 90\% agreed or strongly agreed that it was easy to learn to use the system (Figure~\ref{fig:pssuq}).} According to P7, when executing larger shifts, they don't necessarily have a clear mapping to algorithms and attributes in the code. Instead, natural language is more expressive in communicating their creative intent: \textit{``it matches to my expectations better on some of the conceptually large changes in the sense that I like didn't have to be as precise\ldots''} (P7) Similarly, P11 commented: \textit{``I felt a better immediate response with the larger conceptual changes, just because you could at least get some of the feelings that I was trying to capture. ''} In general, participants responded positively to the generated code outputs. As P6 commented about making a rainbow effect:\textit{``This is exactly what I envisioned when I said, Make a rainbow. So that's pretty cool. That's close.''}

For smaller syntactic variations, participants had mixed opinions about using prompts. For example, P1 found the prompt-based interactions less useful when they already had a clear idea: \textit{``I don't really see this being good for when I have an idea of what I want. It seems way more frustrating to try to chat with it to give me what I want versus coding it myself.''} However, P9 commented about the prompt affordances that \system allowed them to do things they would typically avoid: ``[with Spellburst] I found myself doing things that I wouldn’t normally do because for-loops are kind of a drag for me\ldots but with \system I was like,  Oh, I don’t have to write out all that syntax for these for-loops.'' P2 commented that syntactic prompts could, in fact, make creative coding more accessible to non-coders. These varied experiences suggest that the balance between semantic and syntactic prompting should be carefully considered in designing future systems like \system.

When using prompts to generate code outputs, participants found the AI auto-complete suggestions beneficial in providing ideas. According to P10: ``I did find some inspiration in the suggestion\ldots It took me down some paths that I didn't really think I was gonna go down, and I always appreciate that like opportunity. so I can almost imagine like a little clippy-style AI \ldots here's a weird thing you could try with this output like just giving me ideas for where to go next.'' However, several participants expressed the need for interpretability of the generative outputs. P5 commented:\textit{``I think the main thing I would want is more interpretability, like perhaps [currently] a little bit too much magic."} and \textit{``I find myself wanting to be able to turn each part [of the prompt] on and off to figure out which part is contributing to which."} 

The primary error encountered with Spellburst was related to bugs in the generated code that would either prevent that specific output from running or produce output that was far from the participant's expectations. To address errors associated with AI generated code, we do allow artists to directly access the code editor.  In cases where generated code did not run, the errors were isolated to one path, which participants typically addressed by trying again with a different prompt or going into the code editor. Several participants were eager to spend more time figuring out the generated errors in the code. However, the limited duration of the sessions served as a constraint. Future research could explore the trade-offs associated with using generated AI code that may be error-prone and require time and effort on manual bug-fixing modifications. P10 expressed how errors encountered through prompt generation could disrupt creative flow as it puts them in a ``weird, intermediate space between image and code.''

\subsubsection{Comparison of \system with ChatGPT baseline}
Based on feedback from using both \system and their typical IDE alongside ChatGPT, a majority of artists found the visual interface, sketch organization, and branching features of Spellburst to be key factors that make it preferable when compared to ChatGPT. For example, P1 appreciated the branching capability of \system, which enabled rapid creative exploration in comparison to a more linear process with ChatGPT: \textit{``I think being able to go between the different branches was really interesting to me\ldots modifying on a specific sketch that I see, like seeing the way that was laid out was really cool.''} Similarly, P9 commented how the visual iterations in \system felt like a different process compared to ChatGPT and their regular IDE: 
\begin{quote}
   \textit{ ``\ldots after the experience of using the raw ChatGPT version, I really did come to appreciate the sort of ease with which you could reason about the different versions, and very precisely sort of branch and duplicate and not have it be ambiguous sort of which one you were basic off of and sort of explore. The ChatGPT version felt very linear, whereas this one seemed to very much encourage branching exploration and making that sort of a first-class part of it. So that was cool.''} (P9)
\end{quote}

P8 even expressed how \system could accelerate their creativity: ``I was going faster with the new interface for sure able to explore way more ideas per minute." However, in contrast, P4 mentioned reduced friction when using ChatGPT due to greater familiarity: \textit{``The ChatGPT interface is somewhat known to me. so I feel there is slightly less friction going in there [versus \system].''}

\subsubsection{Usability of \system's Interface}

Immediately following the evaluation study, we asked participants to complete both NASA Task Load Index (NASA-TLX) \cite{hart1988development} and the Post-Study System Usability Questionnaire \cite{lewis1992psychometric}.
In the NASA-TLX survey results, participants found Spellburst to be the most mentally demanding ($mean = 9.9$, $SD = 4.8$). On the other hand, they found it to be the least physically demanding ($mean = 2.6$, $SD = 1.2$). There was notable variation in participants' experiences in terms of success in accomplishing what they were asked to do with Spellburst ($mean = 11.4$, $SD = 5.2$), as well as mental demand ($mean = 9.9$, $SD = 4.8$) and stress with Spellburst ($mean = 7.1$, $SD = 4.4$).

In the PSSUQ survey results, we found the overall usability score for \system as 7.67. In terms of sub-scales, the System Usefulness (SYSUSE) score was 8.26 which indicates that the system was relatively easy to use and learn. The Information Quality (INFOQUAL) score was 6.76, which suggests that while the information provided by the system was helpful, there is from for improvement in areas such as error messages, documentation, and information organization. The Interface Quality (INTERQUAL) score was 8.24, which indicates that participants found the system's interface to be pleasant and visually appealing. Combined, these results indicate that \system has a good level of usability, but some aspects, particularly information quality, could be further improved to enhance the user experience. 

\begin{figure*}[htb]
  \centering
  \includegraphics[width=\textwidth]{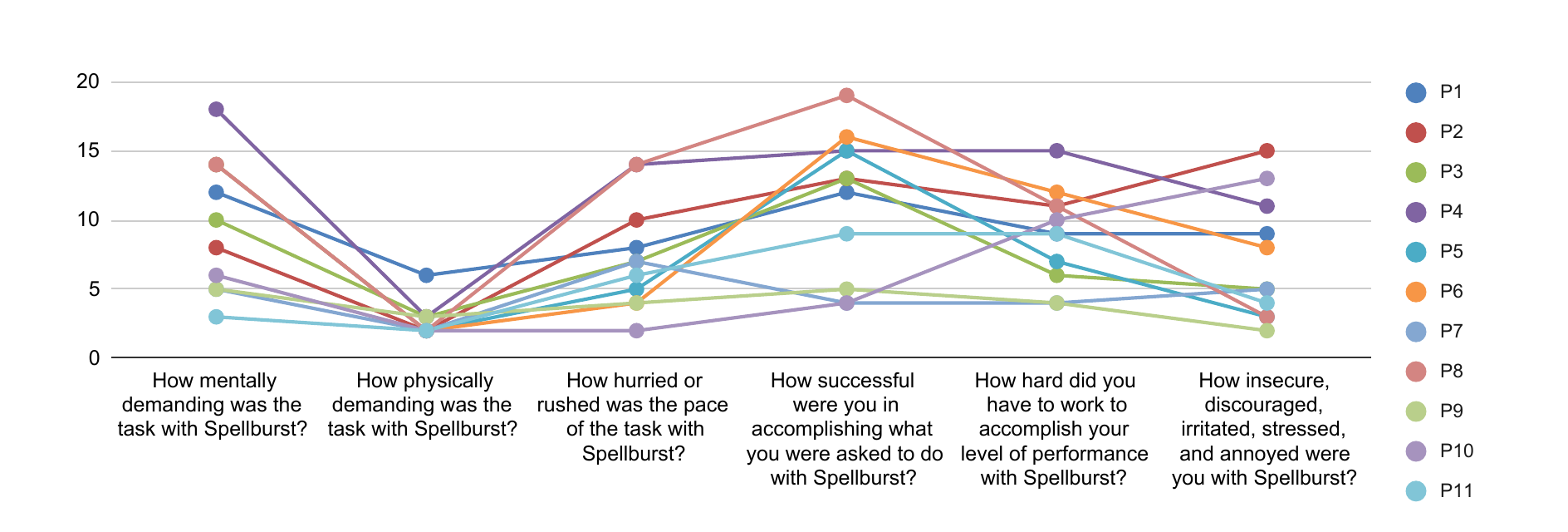}
  \caption{Participant Responses to the NASA-TLX Cognitive Load Questionnaire}
  \label{fig:nasatlx}
\end{figure*}

Participants appreciated Spellburst's visual branching and merging without having to engage with Git-style commands. Many of the participants did not typically engage with Git for version control and appreciated how \system could help seamlessly organize their projects without too much overload, with P7 sharing how it can address their pitfall of focusing too much on organizing: \textit{``I find that anytime I get excited about trying to organize the creative project. All the excitement goes into the organizing.''} Further, P10 appreciated the automatic branching  functionality and the automatic rendering of the code, feeling that \textit{``this would work really well for me across my P5 sketches.''} The visual version control mechanics of \system also served to reduce creative fiction for P8: \textit{``You don't have to be scared that you're going to lose a pretty state.''}

Participants also discussed the potential for Spellburst to impact their workflow as P7 shares: \textit{``Maybe we do have something that's maintained in a Git repository, and [\system] is like a sandbox for me to change small pieces of code and see how it impacts the entire project.''} To spur rapid iteration and creative exploration, P6 took advantage of running multiple prompts simultaneously and received an unanticipated benefit from the delay time associated with AI code generation: \textit{``It means you can start off ideas and not be influenced by the output of the first one when you're thinking about the second one \ldots this gives you time to continue thinking of ways to remix the previous ones without being influenced.''}

However, some expressed a desire for bookmarking nodes or a way to clear unrelated content and focus on a specific branch without visual overload, as expressed by P8: \textit{``I want to just clear the stage and look at this one\ldots otherwise it can be too much mind clutter.''} The ability to zoom in and out at different levels of detail and give the user control over both output and focus space in the UI are considerations for future development.

\subsubsection{Perceptions of impact on their creative practice}

Participants expressed interest in a tool like this to support broader exploration and ideation in earlier stages of projects. Almost all found the tool beneficial for fast prototyping, with P1 likening the experience \textit{``\ldots as a practice like the same way you doodle on a napkin"}. While several participants felt that the generated code typically did not map exactly to their expected output, it was perceived to help with ideation, with \system uniquely designed to support visual navigation across different iterations. P9, who had limited prior experience using prompts to create code, noted how \system encouraged her to \textit{``\ldots reflect about the way I was thinking about code \ldots normally when I teach coding we're giving very specific instructions to the computer. But now I'm giving less specific, more abstracted instructions \ldots and in this case they don't need to be very specific [in order to be successful]."}

However, some concerns were also raised about the responsible and ethical development and use of systems like \system that leverage large language models with data trained from large amounts of information on the internet. \edit{This remains an open area of discussion, and the experts that we interviewed had conflicting views about artists' rights to ownership, ranging from concern about plagiarizing to issues of ownership in art being historically muddy; for example, artists who lead teams of assistants may nominally take ownership of work that is actually much more collaborative. Specifically,} some participants were troubled by the possibility of accidental plagiarism and a desire to give credit to prior artists, with P1 stating \textit{``I use other people's code a lot, but it's always credited, and I try to be very clear about that in my sketches at the top [with comments acknowledging] that I've taken from others"}. Perceptions of this did vary by individual, with P5 sharing that \textit{``I would be more concerned about copying somebody else's visual style and less about copying the exact implementation of it."} P9 expressed concern about 1) bias in training sets \textit{``I'd be curious to see how it responds to race and gender"} 2) potential lack of understanding amidst the layers of abstraction \textit{``\ldots there's less of a feeling that you have a grasp of what's going on."}, and 3) skepticism toward AI tools developed by companies solely driven by profit \textit{``seeing some of them built on top of p5 feels like, Oh man, we were really trying to build a tool embedded around access and inclusion and they’re being wrapped by various layers. I would just question, what’s being prioritized in those tools?"}

\edit{These perceptions tie back to prior research that shows how AI obscures key parts of the creative process that artists would have otherwise wanted to customize or understand~\cite{li2021we} and that large language models can regurgitate copyrighted material without correct attribution \cite{henderson2023foundation}, resulting in lawsuits about artists' work being included as training data for LLMs without their consent \cite{vincent_2023}. Given the ethical and legal risks at hand, future work on \system should require that LLMs provide proper citation of the source. This would also benefit users seeking more granularity and explainability; they would be able to follow the original source and see more inspiration from real artists.}

\edit{As some of our participants noted in the evaluation study, AI models carry inherent biases regarding race, gender, and other social constructs \cite{stochasticparrots}. When artists prompt our system to interpret natural language phrases such as ``make this sketch more feminine'' or ``make this sketch more aggressive,'' social biases embedded in the model may be expressed in code or AI-generated descriptions of the code, as P9 demonstrated in her user study.} \edit{As this field develops, we must attend to and contribute to the discussions around the responsible and ethical use of generative AI. Until then, we recognize and support the agency of each artist to decide whether or not to use AI-augmented CSTs such as \system. }
\section{Discussion}

\subsection{Utility of \system} 
In addition to the workflow for expert creative coders described in Section 4, \system can support a broader range of tasks for creativity and learning. We briefly describe a set of use cases to demonstrate the broader utility of \system.

\subsubsection{Learning Version Control for Creative Coding Novices}
Version control management and creative coding involve techniques and tools that take time to learn. While this study focused primarily on expert usage, many participants commented on the playfulness of \system as an exciting entry point for beginners. \system could be adapted to help novices explore the full design space. This is especially important as novices lack the exposure to semantic constructs that they prefer (patterns, styles, forms) as well as syntactic constructs that they may employ (functions, parameters). As a result, they may get stuck more easily on either side of the divide. For example, a learner who is experimenting with trigonometric functions in p5.js for the first time may not know all possible effects of varying the parameter range. They will also not be able to map a commonly desired effect to the commonly used parameter range of $-\pi/4$ to $\pi/4$ radians. With \system, a learner is free to cycle through many ideas without getting prematurely fixed in one corner of the design space or getting stuck while attempting to produce one desired effect. In addition, \system provides learners with an intuitive interface for version control that allows them to save, reflect on, and debug their creative process.

\subsubsection{Debugging}
The structure of \system as a node-based version control system also lends itself to supporting the practice of debugging for novices and experts alike. For example, imagine a user attempting to debug a complex p5.js sketch for a recursive fractal. If the current sketch is not working, a user can duplicate the existing sketch, inspect the code, and incrementally adjust the code until the bug is resolved. Then, they can merge the code back into the main branch. This can all be done with AI support via the natural language prompt interface. 

\subsubsection{Sharing and Documenting Work}
We learned from our participants that creative coders often share and document their work for others, even when they work alone. Some of our study participants explained that they post screenshots of work-in-progress on social media networks to solicit feedback from peers. Others document their process and code to be disseminated and ``donated'' to a public audience for remixing \cite{shneiderman2002creativity}. Imagine an artist who wants feedback on a new pendulum system they created using \system. They post a link to their own \system instance for everyone to see. A peer opens up the link and finds an intuitive and interactive account of the artist's process. The peer sees the artist's original intent in natural language with each prompt that was used; they also inspect the code that resulted at each step. The peer happens to prefer an earlier iteration rather than the latest sketch; because \system preserves the non-linear nature of the work, the peer is able to easily discover it and point this out to the artist. A beginner who follows this artist on social media also receives the link. They are interested in how this artist created the pendulum; instead of scanning the workflow as the peer did, the beginner traces the artist's process step-by-step to build their understanding of what the artist did.

\subsection{Limitations and Future Work}

While our evaluation study with generative artists provides valuable insights into \system's effectiveness, there are some limitations to consider, all of which open up avenues for future work. 

\edit{Section 6.2 surfaced various limitations of our system related to unusable code, interface unfamiliarity, and cognitive load. Designing effective interfaces for handling generative AI errors remains an important need in future work. Unexpected results were especially common during merge operations. This is partly related to our design choice to favor ease of semantic merging over control: On one hand, the unexpected results led to playful or serendipitous discovery, especially during large creative jumps. On the other, our design limited users' creative control when executing small, precise creative jumps. Future work could consider designs that strike a balance between exploration and control in merge operations, for example, by optionally allowing the user to explicitly specify and manipulate emphasis of aspects of interest from each input.}

\edit{Further, while we made an attempt to demystify the output from various LLMs, our evaluation results indicate that it is hard to predict which prompts would be ``effective'' in generating desired variations. As a result, there is still no perfect mapping from semantic to syntactic changes, which could frustrate users who would otherwise get a lot from the platform. Initial future work could include further fine-tuning the model on examples using our taxonomy. Likewise, we could leverage other work done to make for better prompting experiences, including features to help add constraints to the generation process \cite{isola2018imagetoimage} as well as leveraging the strengths of the chat-based interface to find common ground \cite{gal2022image} \cite{ruiz2022dreambooth}. Future work should also investigate multi-modal prompting that can build referential connections between the visual canvas and text-based prompts.} 
 
\edit{Third, our study only explored a limited set of interactions with \system. The 10 expert creative coders who participated in the evaluation are not representative of the richly diverse population of creative coders and artists. Future studies should involve larger samples with a wider range of backgrounds to better understand the system's performance and utility. In addition, the tasks used in our evaluation did not cover all possible scenarios and use cases that creative coders might encounter. Exploring additional tasks, particularly ``medium-sized'' jumps that are neither too drastic nor too minor, would provide a more comprehensive understanding of \system's affordances at various stages of the creative process (D6). Future work should investigate the effects of long-term, \textit{in-situ} use on creative practice and the extent to which users can adapt to and benefit from the system over time and in the context of real-world artistic workflows. To support more nodes and edges, we may want to consider new approaches to scale the performance and usability of the interface.}

Finally, large-scale creative work often requires an entire studio of artists and producers. For example, consider a game development team that is building a new underwater scene. Two game developers might need to explore many approaches for rendering the currents. They might start from the same sketch and create completely different interactive graphics. When they sit down to discuss, the developers realize that they appreciate the color scheme in one sketch but prefer the physic simulations of the second sketch. Future work should explore such collaborative use of \system.

\section{Conclusion}

In this work, we present \system, a creativity support tool designed to expand creative outputs for artists working with code. The node-based canvas integrates the act of creative exploration and tracking exploration history to reduce the burden on artists and enhance multiple creative strategies, including combinatorial and transformative creativity. \system employs a large language model informed by crowdsourced data to support semantic and syntactic editing and exploration. \system achieves a balance between larger creative shifts using prompts and fine-grained control through dynamic interfaces between prompt and code and direct code editing.  Our evaluation with expert generative artists demonstrates that \system enhances creative practice through rapid exploration while reducing the overhead of managing the exploration process. 

\begin{acks}
We are grateful to the reviewers and our study participants for their time and helpful feedback. We also thank Abdallah AbuHashem, Maxwell Bigman, Professor John Mitchell, Licia He, and Barron Webster for their input during the early iterations of this project. 
\end{acks}

\balance
\bibliographystyle{ACM-Reference-Format}
\bibliography{99_refs}

\clearpage
\appendix
\appendix
\section{Appendix}
\label{sec:appendix}

\subsection{JSON Schemas for Graph Data}
\label{sec:json_schemas}

\edit{Below is an example of the JSON schema for a node.}
\begin{lstlisting}[backgroundcolor = \color{beige}]
    {
      "width": 300,
      "height": 324,
      "id": "wgtt0s",
      "type": "sketch",
      "data": {
        "sourceNode": "root",
        "sourceCode": "\nfunction setup() {\n
        createCanvas(400, 400);\n  background(255);\n
        strokeWeight(2);\n  stroke(0);\n}\n\nfunction draw()
        {\n\n}\n    ",
        "size": {
          "width": 300,
          "height": 300
        }
      },
      "position": {
        "x": 0,
        "y": 0
      },
      "sourcePosition": "right",
      "targetPosition": "left",
      "selected": false,
      "positionAbsolute": {
        "x": 0,
        "y": 0
      }
    }
\end{lstlisting}

\edit{Below is an example of a JSON object for an edge:}
\begin{lstlisting}[backgroundcolor = \color{beige}]
    {
      "id": "wgtt0s=>ic45uc",
      "source": "wgtt0s",
      "target": "ic45uc",
      "type": "connected",
      "selected": false
    }
\end{lstlisting}

\subsection{Prompts for LLMs}
\label{sec:base_prompts}
Below are the prompts we fed into the ChatGPT API to perform each of \system's core operators. 
\edit{The ChatGPT API takes in an array of messages in the request body following this order: \texttt{[System Configuration, Dynamic Context, User Input]}. The \texttt{System Configuration} is a preset preamble for each route based on the task at hand (modification, merging, etc.). The \texttt{Dynamic Context} is a series of \texttt{user} and \texttt{assistant} messages to demonstrate examples of the kinds of output each route should output, where the assistant is primarily tasked with outputting code (rather than responding with natural language chat). The last user example that's sent is formatted in the same way as the previous, generated user messages in the dynamic context, and the final response that the user sees as the sketch is generated is a final \texttt{assistant} output.}

\edit{With these details in mind, we provide the \texttt{System Configuration} and \texttt{Dynamic Context} for reproducibility.}

\subsubsection{Restrictions} \label{sec:restrictions}
For each prompt, we append a set of restrictions. These restrictions helped structure ChatGPT's return values and also addressed common issues that occurred during experimentation. The string of these restrictions is defined as \texttt{@BASE\_RESTRICTIONS}, and is defined in future prompts.

\noindent \texttt{Restrictions: \\ 
- Only respond with code in your output as a raw string. \\ 
- Be as efficient as possible with your implementations. When producing computationally intensive sketches, try to use optimization methods so they run more quickly. \\ 
- If you are ever asked to apply an animation, remember to always remove any calls of the noLoop function to make sure it actually animates. \\
- Comment your code with useful comments. \\
- Remember to be as efficient as possible with your implementations. When producing computationally intensive sketches, try to use optimization methods so they run more quickly. \\}

\subsubsection{Modification} 
The \texttt{System Configuration} prompt is:

\noindent \texttt{You are a creative coding software engineer focused on creating visually stunning graphics, physics simulations, biological simulations, and data visualizations using p5.js. You are excellent at a few things: creating p5.js sketches, modifying p5.js sketches with natural language prompts, and blending multiple sketches together by merging their code in semantically meaningful ways. \\ \\ @BASE\_RESTRICTIONS \\}

\noindent The \texttt{Dynamic Context} is:
\begin{lstlisting}[backgroundcolor = \color{beige}]
const VARIATION_CONTEXT = [
	{
		"role": "user",
		"content": JSON.stringify({
			code: `let x = 100;
let y = 100;
function setup() {
	createCanvas(700, 410);
};
function draw() {
	background(0);
	fill(255);
	rect(x, y, 50, 50);
};
};`,
			variationPrompt: "add a bunch more balls and make them bounce off the bounds"
		})
	},
	{
		"role": "assistant",
		"content": `//${START_TOKEN}
let numCircles = 20;
// Create an empty array to store the circles
let circles = [];
// Set up the canvas and create the circles
function setup() {
  createCanvas(700, 410);
  for (let i = 0; i < numCircles; i++) {
    circles.push({
      // randomly set the x and y coordinates of each circle within the canvas
      x: Math.floor(Math.random() * 700),
      y: Math.floor(Math.random() * 410),
      // set the radius of each circle
      radius: 10,
      // set the x and y velocity of each circle to a random value between 0 and 0.5
      xVel: Math.random() * 0.5,
      yVel: Math.random() * 0.5,
    });
  }
};
function draw() {
  background(0);
  // loop through each circle in the array and move it according to its velocity
  for (let i = 0; i < circles.length; i++) {
    let cir = circles[i];
    cir.x += cir.xVel;
    cir.y += cir.yVel;
    
    // if a circle reaches the edge of the canvas, reverse its direction
    if (cir.x >= width || cir.x <= 0) {
      cir.xVel *= -1;
    }
    if (cir.y >= height || cir.y <= 0) {
      cir.yVel *= -1;
    }
    
    // set the fill color to white and draw the circle at its current position
    fill(255);
    ellipse(cir.x, cir.y, cir.radius);
  }
};
//${END_TOKEN}`
	}
]
\end{lstlisting}

\subsubsection{Merging} 
The \texttt{System Configuration} prompt is:

\noindent \texttt{Given two p5.js code snippets, generate a new code snippet that combines the functionality of both snippets. The output code snippet should be valid p5.js code and should have as much similarity as possible to the original inputs. \\ \\
First, you'll begin your generation by creating a "merge prompt". This can either be supplied by the user, otherwise you will create it. \\ \\
It should follow this format: "Combine [Feature A] from [Code Snippet 1] with [Feature B] from [Code Snippet 2]. The resulting code should [Describe desired functionality]." \\ \\
In this format, you would fill in the placeholders with the relevant information for your specific merge prompt. For example:
\`/*Combine the animation loop from Code Snippet 1 with the mouse-interactivity of Code Snippet 2. The resulting code should draw a looped animation that responds to user mouse movement by changing its direction and speed in real-time.*/\` \\ \\
Then you'll produce the relevant p5 code according to the prompt and the format provided in the following examples. \\ \\
@BASE\_RESTRICTIONS \\
- Remember to include the merge prompt inside of code comments. \\
}

\noindent The \texttt{Dynamic Context} is:
\begin{lstlisting}[backgroundcolor = \color{beige}]
const exampleInput = {
	firstCode: `let angle = 0;
let r = 100;
function setup() {
  createCanvas(400, 400);
  background(220);
}
function draw() {
  translate(width / 2, height / 2);
  rotate(angle);
  strokeWeight(2);
  stroke(0);
  line(0, 0, r, 0);
  angle += 0.05;
}
`,
	secondCode: `let x, y;
let speed = 3;
function setup() {
  createCanvas(400, 400);
  x = width / 2;
  y = height / 2;
}
function draw() {
  background(220);
  ellipse(x, y, 50, 50);
  x += speed;
  if (x > width || x < 0) {
    speed *= -1;
  }
}`
}

const exampleOutput = `//${START_TOKEN}
 /* Combine the rotating line animation from Snippet 1 with the bouncing ball behavior from Snippet 2. The resulting code should draw a rotating line that bounces off the walls of the canvas and leaves a trail of dots or other shapes */
 
let angle = 0;
let r = 100;
let x, y;
let speed = 3;
function setup() {
  createCanvas(400, 400);
  x = width / 2;
  y = height / 2;
  background(220);
}
function draw() {
  translate(width / 2, height / 2);
  rotate(angle);
  strokeWeight(2);
  stroke(0);
  line(r, 0, x - width / 2, y - height / 2);
  angle += 0.05;
  ellipse(x, y, 5, 5);
  x += speed;
  if (x > width || x < 0) {
    speed *= -1;
  }
  y = height / 2 + sin(x * 0.02) * 100;
}
//${END_TOKEN}
`

const MERGE_CONTEXT = [
	{
		role: "user",
		content: JSON.stringify(exampleInput)
	},
	{
		role: "assistant",
		content: exampleOutput
	}
]

\end{lstlisting}

\subsubsection{Autocomplete}
The \texttt{System Configuration} prompt is:

\noindent \texttt{Your role is to provide autocomplete results for a natural language prompt-editor for creating p5.js sketches. \\ \\
If the input prompt is an incomplete sentence, provide results that continue the sentence. If the input prompt is a complete sentence, provide more complete sentences. \\ \\
You will *always* provide results or suggestions, even if the input seems incomplete. \\ \\
Provide maximum 3 (three) suggestion results. Do not respond with any english. You are not a chat. You are simply returning arrays of data. \\}

\noindent The \texttt{Dynamic Context} is:
\begin{lstlisting}[backgroundcolor = \color{beige}]
const AUTOCOMPLETE_CONTEXT = [
	{
		"role": "user",
		"content": "make an intricate tree with branches that twist and turn, gradually tapering off into smaller and smaller branches."
	},
	{
		"role": "assistant",
		"content": "[\"add variation in color and thickness to branches\",\"randomize branch angles and lengths\",\"incorporate falling leaves or flowers\"]"
	},
	{
		"role": "user",
		"content": "make it more"
	},
	{
		"role": "assistant",
		"content": "[\"colorful\", \"sporadic and physical\", \"like a surreal drawing\"]"
	},
	{
		"role": "user",
		"content": "draw numerous small particles"
	},
	{
		"role": "assistant",
		"content": "[\"that are attracted to each other with a gravity well\",\"that respond to user input to change particle behavior\",\"that collide with each other\"]"
	},
	{
		"role": "user",
		"content": "create an abstract and "
	},
	{
		"role": "assistant",
		"content": "[\"visually striking piece of art using perlin noise.\",\"experiment with color gradients and blending modes\",\"incorporate user input for dynamic patterns\"]"
	},
]
\end{lstlisting}

\subsubsection{Extraction}
The \texttt{System Configuration} prompt is:

\noindent \texttt{You are the most experienced creative coding assistant in the world who is focused on creating visually stunning graphics, physics simulations, biological simulations, and data visualizations using p5.js. You can help answer coding questions, write code, and change code. Specifically, you are excellent at answering questions about p5.js sketches. \\ \\
You have read countless articles about building interactive art and graphics, and have read everything from the p5.js API documentation (https://p5js.org/reference/), as well as all of the "Nature of Code" articles and tutorials (https://natureofcode.com/book/). \\}

\subsubsection{Diffing}
The \texttt{System Configuration} prompt is:
\noindent \texttt{You are the most experienced creative coding assistant in the world who is focused on creating visually stunning graphics, physics simulations, biological simulations, and data visualizations using p5.js. \\ \\
You have read countless articles about building interactive art and graphics, and have read everything from the p5.js API documentation (https://p5js.org/reference/), as well as all of the "Nature of Code" articles and tutorials (https://natureofcode.com/book/). \\ \\
Compare the two pieces of p5.js code. In no more than 5 sentences, describe how they are similar and different. Focus on the content of each sketch, their properties (such as color and stroke), and code-level differences. Don't propose a function. \\}

\subsection{Crowd-sourcing Image Transformations}
\label{sec:image_transformation}

\subsubsection{Design of Survey}
After developing our taxonomy, coding the examples, and selecting the top $20$ representative examples, we recruited participants from the Prolific platform \cite{palan2018prolific}, targeting adult (18+), English first language speakers with sufficient art experience from the USA and with access to a desktop computer. The initial set of age, location, and language constraints narrowed the eligible pool of participants from roughly ~120,000 to ~37,000. With this participant pool in mind, we implemented a pre-screening questionnaire in Prolific to assess the participants' art experience and self-perceptions in relation to art. The questions were inspired from a gifted and talented art screening questionnaire \cite{bangorschool}. A total of 370 participants completed the initial pre-screen survey.

We then invited the eligible participants to complete a series of crowdsourced tasks, each of which involved them providing a natural language prompt to transform an image from one state to the next. The tasks were designed to only elicit participants' natural language descriptions, without exposing them to the underlying code.

\begin{figure}[htb]
  \centering
  \includegraphics[width=\columnwidth]{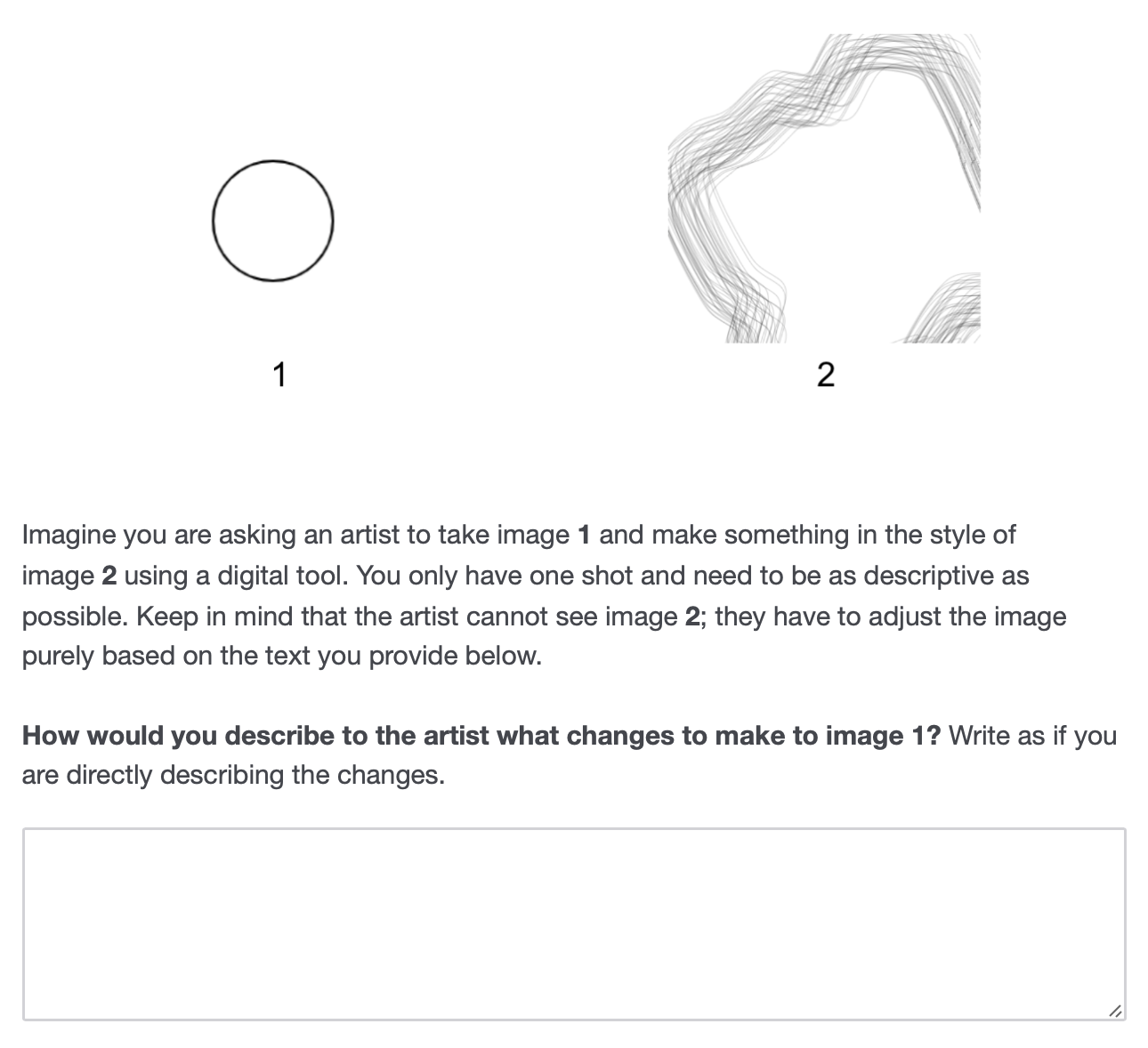}
  \caption{Example crowdsourced task}
  \label{fig:crowdsourcedtaskexample}
\end{figure}

Each task included an initial starter image output of a creative coding project and a final image (see Figure \ref{fig:crowdsourcedtaskexample} for an example). Participants were asked to provide a prompt that they believed would sufficiently describe the transformation from the starter image to the final image. Incomplete and timed out crowdsourced tasks were flagged by Prolific and discarded. In total, we collected 12 prompts for each of the 20 images with input from 52 unique participants on at least one prompt.

We categorized the crowd-sourced responses into three broad categories: semantic descriptions (``make it look like an iceberg melting""), syntactic descriptions (``increase the stroke weight of the lines''), and \textit{combined semantic and syntactic descriptions} (``randomly vary the thickness like tree branches''). We found that a common pattern for successful \textit{combined semantic and syntactic descriptions} involved reference to a taxonomy term as well as reference to a simile or metaphor.

\subsubsection{Example Prompts}
Here are some example prompts from the survey:
\begin{itemize}
    \item \textit{With the circle image, create a finale image that has many circles echoing and rippling out all around it. As the shapes ripples out, the shades fade out in greyscale until they're almost invisible.  They distorted and change shape and are no longer perfect circles. To achieve this, lower the images levels by 30\%. Then multiple the original circle layer and lower the levels 5\% more on the new layer. With the new circle layer than distort and manipulate it's shape. Repeat these steps again, and again, until dozes of distorted shapes are echoing out from the original circle layer.}
    \item \textit{Take the different sides of the line of the provided image (circle) and use the tool to stretch the line in and out making a sort of wave effect around the whole circle. Then copy this line over itself many times at different points either farther inside the original circle line or farther outside the original circle line.}
    \item \textit{instead of a circle, it would look like multiple pencil lines overlapping to create an organic looking shape.} 
    \item \textit{You will need to add a bunch of lines to image 1. You will need to draw an irregular circle multiple times.}
    \item \textit{I would open it up so that it is no longer a circle and make it look more vibrant and less fluid. It will have a distressed quality and will look a little frayed. It will have blurred lines that will overlap one another. It won't be a closed circle and will be a lot larger. The blackness will also be a little more like a grey color.}
\end{itemize}

\subsection{Example \system Evaluation Output}
\label{sec:example_output}

\begin{figure*}[htb]
  \centering
  \includegraphics[width=\textwidth]{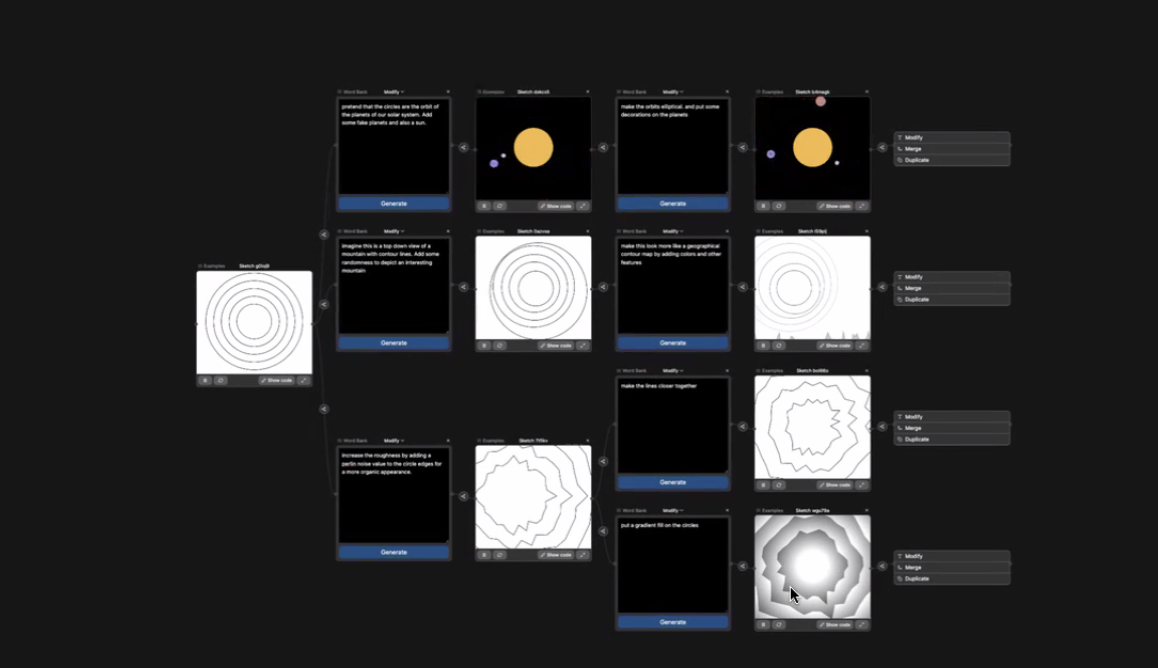}
  \caption{Evaluation study sketch from P6 indicating encompassing several variations on concentric circles starter code.}
  \label{fig:sketch1}
\end{figure*}

\begin{figure*}[htb]
  \centering
  \includegraphics[width=\textwidth]{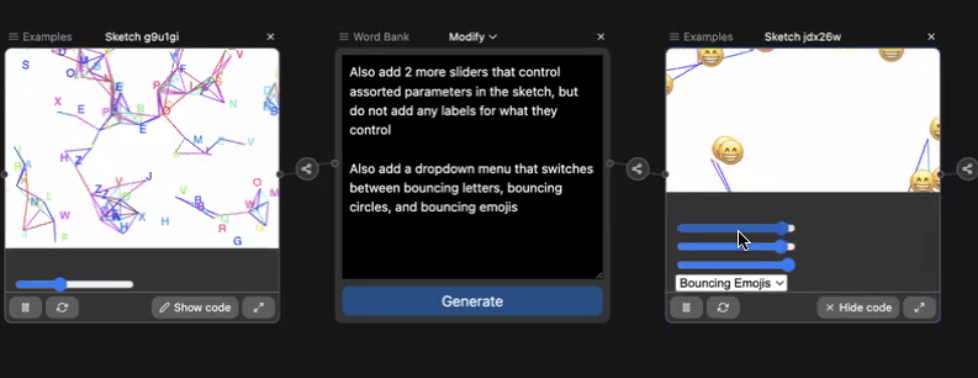}
  \caption{Evaluation study sketch from P8 indicating the user's generation of a provisional UI in the form of sliders to support their creative exploration.}
  \label{fig:sketch2}
\end{figure*}

\newpage
\subsection{System Architecture}
\label{sec:architecture_diagram}
\begin{figure*}[htbp]
  \centering
  \includegraphics[width=\textwidth]{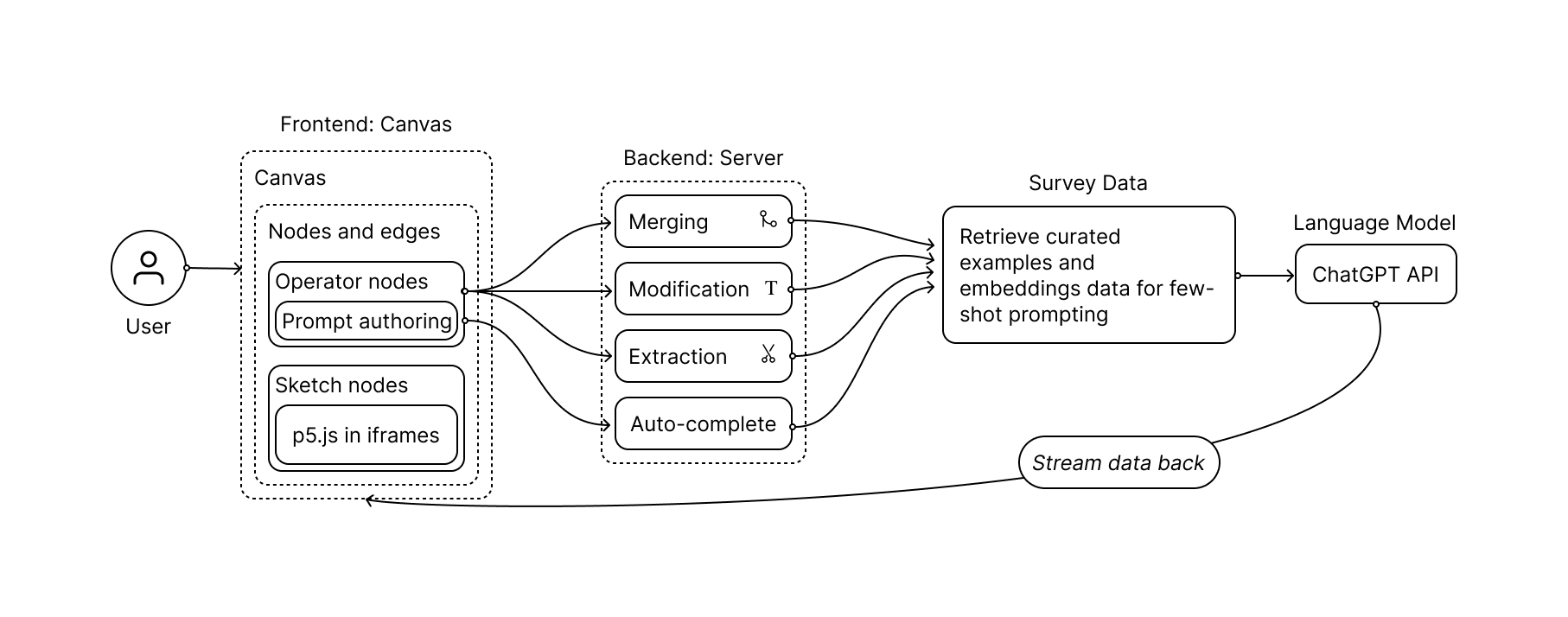}
  \caption{System architecture of Spellburst, showing the connection between the front-end Canvas and the server that calls the ChatGPT API.}
  \label{fig:architecture}
\end{figure*}

\newpage
\subsection{Demographic Information of Participants}
\label{sec:demographic_info}

\begin{table}[H]
    \centering
    \begin{tabular}{|c|c|c|c|c|}
     \hline
     & Gender & Experience & Background & Tools \\ 
     & & (Years) & & \\
     \hline
     U1 & M & 5+ & Creative Coding & p5.js, Python \\
     & & & Educator & \\
     \hline 
     U2 & F & 5+ & Generative Artist & Javascript, \\
     & & & & Python \\
     \hline 
     U3 & F & 5+ & Generative Artist & Javascript \\
     & & & & p5.js \\
     \hline 
     U4 & M & 3-5 & Visual Designer & Adobe, Figma \\
     & & & & React.js \\
     \hline 
     U5 & M & 5+ & Software Engineer & Javascript \\
     \hline 
     U6 & M & 5+ & Software Engineer & Javascript, \\
     & & & & HTML, CSS \\
     \hline 
     U7 & M & 5+ & Game Developer & Unreal \\
     & & & & Engine \\
     \hline 
     U8 & F & 5+ & Visual Designer & Figma, \\
     & & & & SVG Art \\
     \hline 
     U9 & M & 5+ & Generative Artist & Javascript \\
     \hline 
     U10 & M & 5+ & Generative Artist & Javascript \\
     \hline 
    \end{tabular}
    \caption{Generative Study Participants Demographics}
    \label{tab:generative_study}
\end{table}


\begin{table}[H]
    \centering
    \begin{tabular}{|c|c|c|c|c|}
     \hline
     & Gender & Experience & Background & Tools \\ 
     & & (Years) &  & \\
     \hline
     P1 & F & 2.5 & Generative Artist & C++, \\
     & & & & Javascript \\
     \hline 
     P2 & M & 10-11 & UI Designer, & HTML, CSS \\
     & & & Creative Coder & Javascript \\
     \hline 
     P3 & M & 10-15 & Software Engineer & p5.js, \\
     & & & & 3D Modeling \\
     \hline 
     P4 & M & 4 & HCI Researcher & Sketch-n- \\
     & & & & Sketch \\
     \hline 
     P5 & M & 6 & Creative Coder & LED and \\
     & & & & Video Art \\
     \hline 
     P6 & M & 5-6 & Generative Artist, & HTML, CSS, \\
     & & & Software Engineer & p5.js \\
     \hline 
     P7 & M & 8-10 & Digital Artist, & Python, \\
     & & & Data Scientist & Blender \\
     \hline 
     P8 & M & 5 & Creative Coder, & Processing, \\
     & & & College Student & Python \\
     \hline 
     P9 & F & 15+ & Creative Coder, & Swift, Java, \\
     & & & Open Source Creator & HTML, CSS \\
     \hline 
     P10 & M & 10 & Visual Designer & HTML, CSS, \\
     & & & & 3D Modeling \\
     \hline 
     P11 & F & 4 & Creative Coder & p5.js, \\
     & & & & Unity (C\#) \\
     \hline 
    \end{tabular}
    \caption{Expert Evaluation Participants Demographics}
    \label{tab:expert_evaluation}
\end{table}

\end{document}